\documentclass[aip,apl,reprint,dvipdfmx]{revtex4-1}

\usepackage{graphicx}
\usepackage{amsmath}
\usepackage{dcolumn}
\usepackage[bookmarks=false,colorlinks=true,linkcolor=blue,citecolor=blue,urlcolor=blue]{hyperref}
\usepackage{xr}
\usepackage{here}
\usepackage{bm}
\usepackage{framed}

\begin{document}

\title{Development of the temperature-dependent interatomic potential for molecular dynamics simulation of metal irradiated with an ultrashort pulse laser}
\author{Yuta Tanaka}
\email[Electronic mail: ]{tanaka@cms.phys.s.u-tokyo.ac.jp}  
\email[Present address: Nippon Steel Corporation, 20-1 Shintomi, Futtsu, Chiba 293-8511, Japan; ]{tanaka.968.yuuta@jp.nipponsteel.com}
\affiliation{Department of Physics, The University of Tokyo, 7-3-1 Hongo, Bunkyo-ku, Tokyo 113-0033, Japan}
\author{Shinji Tsuneyuki}
\affiliation{Department of Physics, The University of Tokyo, 7-3-1 Hongo, Bunkyo-ku, Tokyo 113-0033, Japan}
\affiliation{Institute for Solid State Physics, The University of Tokyo, 5-1-5 Kashiwanoha, Kashiwa, Chiba 277-8581, Japan}

\date{\today}

\begin{abstract}

Laser ablation is often explained by a two-temperature model (TTM) with different electron and lattice temperatures. To realize a classical molecular dynamics simulation of the TTM, we propose an extension of the embedded atom method to construct an interatomic potential that is dependent on the electron temperature. This method is applied to copper, and its validity is demonstrated by comparison of several physical properties, such as the energy-volume curve, phonon dispersion, electronic heat capacity, ablation threshold, and mean square displacement of atoms, with those of finite-temperature density functional theory.

 \end{abstract}


\maketitle

\section{Introduction}
\label{sec:intro}

The development of ultrashort pulse lasers has opened up a new research field, where many intensive investigations have been performed due to the peculiarity of the phenomena caused by irradiation with ultrashort pulse lasers.
Irradiation of ultrashort pulse laser causes a nonequilibrium state that is extremely different from the equilibrium state.
Specific phenomena observed in the nonequilibrium state have been reported to date.
For example, irradiation of a ultrashort pulse laser onto a solid surface produces an ultrafast structural change,~\cite{Fritz_2007,Daraszewicz_2013} coherent phonon,~\cite{Hase_2009} hot plasma confined inside a cold solid, and the emission of excessively high-energy atoms/ions~\cite{Miyasaka_2012,Hashida_2010,Dachraoui_2006,Dachraoui_2006_2} have been reported.

The nonequilibrium state is very complicated, so that no systematic description has been offered to date.
One of the most employed methods to investigate ultrashort laser-irradiated metals is the well-known two-temperature model (TTM).~\cite{Anishimov_1973}
Ultrashort pulse laser irradiation on a metal surface changes the electron subsystem from the ground state into excited states by absorption of single or higher order multi-photons.
The electron subsystem is thermalized to the Fermi-Dirac distribution with the electron temperature $T_e$, via electron-electron interaction, of which the scattering time $\tau_{ee}$ is approximately $10 \mathchar`-100\,\text{fs}$ in metals.~\cite{Mueller_2013, Brown_2016_2}
In this time scale, the electron subsystem and the lattice subsystem have not reached the local equilibrium state, so that $T_e$ is higher than the lattice  temperature, $T_l$.
The maximum $T_e$ typically reaches more than 10 times higher than the final temperature ($T_e \approx T_l$) because the heat capacity of electrons is excessively smaller than that of the lattice.
$T_l$ begins to increase by energy transfer from the electron subsystem via electron-phonon scattering, of which the relaxation time $\tau_{el}$ is larger than several picoseconds.~\cite{Schoenlein_1987,Elsayed-Ali_1987,Elsayed-Ali_1991,Hohlfeld_2000}
Therefore, under the assumption of instantaneous and local thermalization in the electron subsystem and the lattice subsystem,
ultrashort laser-irradiated metals can be described as $T_e \gg T_l$, long before $\tau_{el}$.
This explanation is the main concept of the TTM.

Based on this TTM with the finite-temperature density functional theory (FTDFT), previous simulations have succeeded in reproducing the phenomena of laser irradiated materials, such as the resolution of the Jahn-Teller distortion in bismuth~\cite{Fritz_2007} and bond hardening in gold.~\cite{Ernstorfer_2009}
It has been implied that these phenomena are caused due to the change in the force acting on atoms by the effect of not only modified Coulomb interaction, but also by the electronic entropy effect at high $T_e$.
Furthermore, it has been proposed that the electron entropy effect leads to lattice instability, which leads to laser ablation.~\cite{Tanaka_2018}

Molecular dynamics (MD) simulations using the $T_e$-dependent interatomic potential (IAP) have successfully represented experimental results at relatively low $T_e$, such as the ultrafast melting of gold.~\cite{Daraszewicz_2013}
However, the validity of the $T_e$-dependent IAP for phenomena at high $T_e$, where laser ablation occurs, has been under debate.
For MD simulations of laser ablation, the $T_e$-dependent IAP has been extended based on the embedded atom model (EAM) potential.~\cite{Norman_2012,Norman_2013,Khakshouri_2008}
Norman~$et\,al.$~\cite{Norman_2012,Norman_2013} used the conventional EAM function form and extended it to finite $T_e$ by making the fitting parameters dependent on $T_e$.
The parameters are fitted to the FTDFT results at several discrete $T_e$, and the values of parameters at temperatures without fitting are interpolated.
In this approach, the force calculated by the IAP is largely underestimated compared with that calculated by FTDFT in the high $T_e$ region ($> 2\,\text{eV}$).
This result indicates that it is not clear whether the interatomic force at finite temperature can be fitted by the conventional EAM function.
Khakshouri~$et\,al.$~\cite{Khakshouri_2008} extended the EAM potential to finite temperatures using the approximation that the electron distribution at finite temperatures changes according to the Fermi-Dirac distribution.
Furthermore, they assumed that the density of states (DOS) of $s$-$p$ bands above $d$ bands are the same as that of $d$ bands, and the DOS of the $s$-$p$ bands continue to infinity. 
They obtained an exact solution of the $T_e$-dependent EAM function form using this simplification.
Although a certification of this IAP is verified around the equilibrium volume, its validity with respect to simulations with large displacement from the equilibrium position and large volume change has not been sufficiently investigated.

Care should be taken in simulations of ablation caused by ultrashort pulse laser irradiation because significantly large volume changes and large displacement from the equilibrium position are involved in this phenomena.
In addition, the dynamics around the equilibrium volume, such as the phonon dispersion and bulk modulus, are implied to be important for representation of the spallation processes during ablation.~\cite{Wu_2013}
Therefore, for simulation of ablation using the IAP, not only the transferability with respect to the volume and atom positions, but also the atom dynamics near equilibrium are important.
However, a $T_e$-dependent IAP has not yet been developed to be able to deal with such problems.

In this paper, we propose a function of the $T_e$-dependent IAP within the EAM potential in Sec.~\ref{sec:IAP}.
We then suggest a parameter fitting strategy for the $T_e$-dependent IAP parameters in Sec.~\ref{sec:fitting}.
From a comparison with the calculation results based on FTDFT, we also show that the $T_e$-dependent IAP reproduces important physical properties for ultrashort pulse laser ablation, such as the mean square displacement and the ablation threshold $T_e$, in Sec.~\ref{sec:result}.
Finally, we provide conclusions in Sec.~\ref{sec:connclusion}.

\section{Electron-temperature-dependent interatomic potential ($T_e$-dependent IAP)}
\label{sec:IAP}

\subsection{Embedded atom method (EAM) potential ($T_e = 0$)}
\label{sec:TeIAP}

We start from a brief review of the conventional EAM potential at $T_e=0\,\text{K}$.
The EAM potential~\cite{Daw_1983} is a simple empirical many-body potential for metals.
In this potential, the total energy of a system $ E_{\text{tot}}$ is expressed as
\begin{eqnarray}
  E_{\text{tot}} = E_{\text{two}} + E_{\text{emb}}.  \label{eq:EAM}
\end{eqnarray}
Here, $E_{\text{two}}$ is the two-body potential, and $E_{\text{emb}}$ is the many-body potential or so-called embedded potential.
$E_{\text{two}}$ is the repulsive part of the potential energy, and $E_{\text{emb}}$ is the attractive part because it represents the cohesive energy.
$E_{\text{two}}$ is generally written as
\begin{eqnarray}
  E_{\text{two}} =  \frac{1}{2} \sum_{i}^N \sum_{j \ne i}^N V(r_{ij}),  \label{eq:two-body}
\end{eqnarray}
where $i$ and $j$ are atom indices, $r_{ij}$ represents the distance between the $i$-th and the $j$-th atoms, $V$ is the pair function, and $N$ is the total number of atoms. 
Finnis and Sinclair~\cite{Finnis_1984} proposed the following form as the function form of $E_{\text{emb}}$:
\begin{eqnarray}
  E_{\text{emb}} =  -A \sum_i^N \sqrt{{\rho}_i},  \label{eq:FS}
\end{eqnarray}
where $A$ is a fitting parameter, and
\begin{eqnarray}
  {\rho}_i =  \sum_{j \ne i}^N {\phi}(r_{ij}),  \label{eq:rho}
\end{eqnarray}
where the function $\phi (r_{ij})$ is the pair potential and is dependent on only $r_{ij}$.
 $\rho _i $ represents the host electron density at the $i$-th atom created by its surrounding atoms.

From here, we derive Eqs.~(\ref{eq:FS}) and~(\ref{eq:rho}) under the rectangular model. 
The electronic states at the $i$-th atom are expected to be described by the local density of states $d_i(E)$.
The band energy $E_i^\text{band}$ of the $i$-th atom can be expressed as~\cite{Sutton} 
\begin{eqnarray}
  E_i^{\text{band}} =  2\int^{\infty}_{-\infty} f(E)(E-E_i^c)  d_i (E) dE,  \label{eq:Eband}
\end{eqnarray}
where $E_i^c$ is the center energy of $d_i(E)$ and $f (E)$ is the Fermi-Dirac distribution.
Using $E_i^{\text{band}}$, $E_{\text{emb}}$ is expressed as
\begin{eqnarray}
  E_i^{\text{emb}} =  \sum_i^{N}  E_i^{\text{band}}.  \label{eq:sumEband}
\end{eqnarray}
In the case of $T_e = 0$, this equation can be written as
\begin{eqnarray}
  E_i^{\text{band}} =  2\int^{E_F}_{-\infty}  (E-E_i^c)d_i (E) dE,  \label{eq:Eband2}
\end{eqnarray}
where $E_F$ is the Fermi energy.
Under the rectangular model, $d_i(E)$ can be written as
\begin{align}
d_i (E) = \begin{cases}
                   \frac{2N^a}{W_i} & \cdot \cdot \cdot  \: E_i^c - \frac{W_i}{2} < E < E_i^c + \frac{W_i}{2} \\
                  \  0   & \cdot \cdot \cdot  \hspace{1.5cm}  \text{the others},
         \end {cases}
\end{align}
where $N^a$ is the number of electron states and $W_i$ represents the band width.
In this model, Eq.~(\ref{eq:Eband2}) is calculated as
\begin{align}
  E_i^{\text{band}} & =  \frac{2N^a}{W_i}   \int^{E_F}_{E_i^c - W_i /2} (E-E_i^c)  dE    \notag  \\
                             & = \frac{N^a}{W_i} \left[ \left( E_F - E_i^c \right) ^2 - \left( \frac{W_i}{2} \right)^2  \right].  \label{eq:Ebandrect2}
\end{align}
Using the number of electrons at the $i$-th atom $N_i^e$, the following equation can be derived:
\begin{eqnarray}
   E_F - E_i^c = \frac{W_i}{2} \left( \frac{N_i^e}{N^a} -1 \right).  \label{eq:Nconserve}
\end{eqnarray}
Using this equation, Eq.~(\ref{eq:Ebandrect2}) can be expressed as
\begin{eqnarray}
   E_i^{\text{band}}  = \frac{N_i^e}{2} \left( \frac{N_i^e}{2N^a} - 1 \right) W_i.     \label{eq:Ebandrect2.5}
\end{eqnarray}
If we assume that the number of electrons at the $i$-th atom is conserved, which is physically reasonable in metals, then $N_i^e$ is a constant value.
$N^a$ is also a constant value so that the following relation can be derived:
\begin{eqnarray}
   E_i^{\text{band}}  \propto  W_i.   \label{eq:Ebandrect3}
\end{eqnarray}

The second moment $\mu^{(2)}_i$, can be calculated as follows
\begin{align}
    \mu^{(2)}_i  & =  \frac{2N^a}{W_i} \int^{E_F}_{E_i^c - \frac{W_i}{2}}  (E-E_i^c)^2 dE \notag  \\
                       & = \frac{1}{12} W_i^2  \left(  \left( \frac{N^e_i}{N^a}-1 \right)^3  + 1 \right)  \notag \\
                       & \propto W_i^2.    \label{eq:EW}   
\end{align}
In the second equality, Eq.~(\ref{eq:Nconserve}) is used.
$\mu^{(2)}_i$ is a summation of the square of the transfer energy $t{(r_{ij})}$, between the atomic orbitals of the $i$-th and $j$-th atoms, so that $\mu^{(2)}_i$ can be expressed as a function of $r_{ij}$.
Combined with Eq.~(\ref{eq:rho}), $\mu^{(2)}_i$ can be written as 
 \begin{eqnarray}
     \mu^{(2)}_i   \propto  \rho _i  \propto \sum_{j \ne i}^N {\phi}(r_{i j}).   \label{eq:Emu2}
\end{eqnarray}
Using Eqs.~(\ref{eq:Ebandrect3}), (\ref{eq:EW}), and (\ref{eq:Emu2}), the following relation can be obtained:
 \begin{eqnarray}
   E_i^\text{band} \propto  W_i   \propto  \sqrt{  \mu^{(2)}_i } \propto  \sqrt{ \rho _i } \propto \sqrt{  \left(   \sum_{j \ne i}^N {\phi}(r_{ij}) \right) }.   \label{eq:Emu}
\end{eqnarray}
Based on this consideration, Eqs.~(\ref{eq:FS}) and~(\ref{eq:rho}) are certified in the case of the rectangular model.

\subsection{Extension to finite $T_e$}
\label{sec:TeIAP}

To construct the function of the $T_e$-dependent IAP that can be applied to the condensed state and the atomic state, we consider the following.

The following function is used as a function form of the $T_e$-dependent IAP for the free energy of the system:
\begin{eqnarray}
  F_{\text{tot}}(T_e) = E_{\text{tot}}(T_e) -S_{\text{tot}}(T_e)T_e,  \label{eq:fp1}
\end{eqnarray}
where $F_{\text{tot}}(T_e)$ and $S_{\text{tot}}(T_e)$ are the total free energy of the system and the total electronic entropy of the system, respectively.
$E_{\text{tot}}(T_e)$ and $S_{\text{tot}}(T_e)$ are expressed by the following equations:
\begin{align}
  E_{\text{tot}}(T_e)& =  E_{\text{two}}(T_e) + E_{\text{emb}}(T_e),  \label{eq:fp2} \\
        S_{\text{tot}}(T_e)    & =   S_{\text{emb}}(T_e)  \label{eq:fp3}.
\end{align}
Here,  $E_{\text{two}}(T_e)$ is the two-body potential, and $E_{\text{emb}}(T_e)$ and $S_{\text{emb}}(T_e)$ are the embedded potentials.
We assume that $S_{\text{tot}}(T_e)$ can be expressed by the embedded potential form because $S_{\text{tot}}(T_e)$ is not attributed to two-body effects, but is attributed to the many-body effect  from the occupation rate of the density of states.
The electronic entropy of the $i$-th atom for independent particles that occupy single-particle states $S_i (T_e) $, is expressed as
\begin{align}
    S_i (T_e) = - 2 k_B  \int^{\infty}_{-\infty}  [f(E)\ln f(E) + (1- f(E)) &   \notag \\
     \times  \ln (1- f(E))]  d_i(E) & dE.  \label{eq:Entropy} 
\end{align}
The total electronic entropy $S_{\text{tot}}(T_e)$ can then be expressed as
\begin{eqnarray}
  S_{\text{emb}} (T_e) =  \sum_i^{N}  S_i  (T_e).  \label{eq:sumSband}
\end{eqnarray}
 
\begin{figure}[tb]
  \begin{center}
    \begin{tabular}{cc}

      \begin{minipage}{1.0\hsize}
        \begin{center}
          \includegraphics[clip, width=9cm]{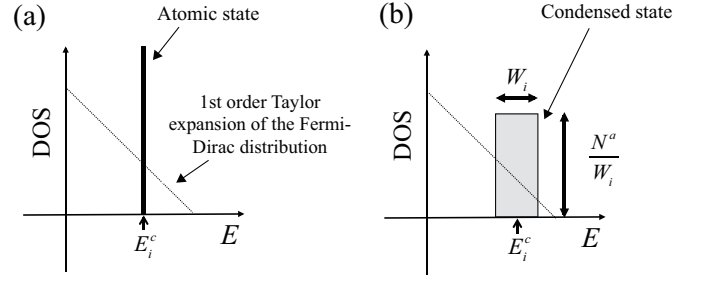}
       
        \end{center}
      \end{minipage} \\
	
   \end{tabular}
    \caption{Schematic image of the consideration. 
    Here, the rectangular model is used.
    (a) The local DOS of the atomic state, and (b) the local DOS of the condensed states.
              }
    \label{fig:band_atom}
  \end{center}
\end{figure} 

Figure~\ref{fig:band_atom} shows a schematic image of the consideration to determine the function form of $E_{\text{emb}}(T_e)$.
The rectangular model is used in this consideration.
In addition, the high $T_e$ limit ($k_BT_e \gg W_i$) is assumed.
Using the first-order Taylor expansion, $f(E)$ can be approximated as 
\begin{eqnarray}
f(E) \cong  \frac{1}{2} - \frac{1}{4k_BT_e} (E- \mu), \label{eq:Fermitailor}
\end{eqnarray}
where $\mu$ represents the chemical potential.
Using Eqs.~(\ref{eq:Eband}) and (\ref{eq:Fermitailor}) under the rectangular model, the following relation can be derived:
\begin{align}
	E_i^{\text{band}} & = \frac{2 N^a}{W_i}  \int^{E_i^c + \frac{1}{2}W_i}_{E_i^c - \frac{1}{2}W_i}    \left[ \frac{1}{2} -  \frac{E - \mu}{4k_BT_e}  \right] (E-E^c_i)   \notag \\
	  & = \frac{2 N^a}{W_i}  \int^{E_i^c + \frac{1}{2}W_i}_{E_i^c - \frac{1}{2}W_i}  \Big[ ( \frac{1}{2} + \frac{\mu}{4k_BT_e} ) E^c_i  \notag \\
	  & \qquad \qquad \qquad + ( \frac{1}{2} +  \frac{ \mu + E_i^c }{4k_BT_e} ) E -  \frac{1}{4k_BT_e} E^2 \Big]   \notag \\
	  & = 2 N^a \Big[- ( \frac{1}{2} + \frac{\mu}{4k_BT_e} ) E^c_i  + ( \frac{1}{2} +  \frac{ \mu + E_i^c }{4k_BT_e} ) E^c_i    \notag  \\
          & \qquad \qquad \qquad - \frac{W_i}{12k_BT_e} \{(3E_i^c)^2 +\frac{1}{4}W_i^2 \} \Big]   \notag \\
	  & = - \frac{N^a W_i^2}{24k_BT_e}  \propto W_i^2.  \label{eq:EcondTe}
 \end{align}
 This relation indicates that if Eq.~(\ref{eq:EW}) is satisfied at high $T_e$, then $E_{\text{emb}}(T_e)$ can be written as a function of $W_i^2$.

The function form of $S_{\text{emb}}(T_e)$ is then derived.
The electronic entropy of the $i$-th atom $S_i (T_e)$, can be written as Eq.~(\ref{eq:Entropy}).
Under the rectangular model and the high $T_e$ limit, this equation can then be expressed as
\begin{align}
    S_i (T_e) = - 2 k_B \frac{N^a}{W_i} \int^{E_i^c + \frac{1}{2}W_i}_{E_i^c - \frac{1}{2}W_i}   [f(E)\ln f(E) & \notag \\
          +  (1- f(E)) \ln (1- f(E))] & dE.  \label{eq:Entropy2} 
\end{align}
Furthermore, using the 2nd-order Taylor expansion of the Fermi-Dirac distribution, this equation is rewritten as
\begin{equation}
     S_i  (T_e)  \cong - 2 k_B  \frac{ N^a}{W_i}\int^{E_i^c + \frac{1}{2}W_i}_{E_i^c - \frac{1}{2}W_i}  \left[   \ln{2} +  \frac{(E- \mu) ^2}{8(k_BT_e)^2}  \right] dE. \label{eq:Entroptailor}
\end{equation}
Using this equation, the electronic entropy of the atomic state $S^{\text{atom}}$, where the local DOS is expressed as the delta function, can be expressed as
\begin{eqnarray}
     S_i^{\text{atom}}  (T_e)  \cong -2 k_B N^a \left[   \ln{2} +  \frac{(E_i^c-\mu) ^2}{8(k_BT_e)^2}  \right]. \label{eq:Entropdelta}
\end{eqnarray}
Similarly, from Eq.~(\ref{eq:Entroptailor}), the electronic entropy of the condensed state $S^{\text{cond}}$, is calculated as 
\begin{align}
     S_i^{\text{cond}}  (T_e)  \cong - 2 k_B N^a  \left[  \ln{2} + \frac{(E_i^c-\mu) ^2}{8(k_BT_e)^2}  \right] & \notag \\
                    +  \frac{W_i^2}{48(k_BT_e)^2}&. \label{eq:Entropband}
\end{align}
Therefore, the following relation can be derived:
\begin{align}
   S_{\text{tot}}(T_e)  & =   S_i^{\text{cond}} - S_i^{\text{atom}}     \notag \\
                      &=   \frac{W_i^2}{48(k_BT_e)^2}   \notag \\
                      & \propto   W_i^2. \label{eq:Entropdelta}
\end{align}

\section{Parameter fitting}
\label{sec:fitting}

\subsection{Fitting parameters}

Here, we explain the fitting parameters to represent the $T_e$-dependent IAP for the free energy in Eq.~(\ref{eq:fp1}).
Using Eqs.~(\ref{eq:EW}), (\ref{eq:Emu2}) and (\ref{eq:EcondTe}), $ E_{\text{emb}}(T_e)$ in Eq.~(\ref{eq:fp2}) can be expressed as
\begin{align}
    E_{\text{emb}}(T_e) & = \alpha_1(T_e) \sum_i^N W_i (T_e) + \alpha_2(T_e) \sum_i^N W_i^2(T_e)  \notag \\
                                 & =  a_1(T_e)  \sum_i^N \sqrt{ \rho _i (T_e) }+  a_2(T_e)  \sum_i^N \rho _i,  
 \end{align}
where $\alpha_1(T_e), \alpha_2(T_e), a_1(T_e)$ and $ a_2(T_e) $ are the fitting parameters, and $ \rho _i $ is described as:
\begin{equation}
    \rho _i  = \sum_{j \ne i}^N \phi (r_{ij}).   \label{eq:fp4}
\end{equation}
The internal energy in Eq.~(\ref{eq:fp2}) is represented as
\begin{align}
  E_{\text{tot}}(T_e) & =  E_{\text{two}}(T_e) + E_{\text{emb}}(T_e)  \notag \\
                               & = \frac{1}{2}  \sum_i^N  \sum_{j \ne i}^N V(r_{ij}) + a_0 (T_e)    \notag \\
                               &  + a_1(T_e) \sum_i^N  \sqrt{ \rho _i (T_e) } + a_2(T_e) \sum_i^N  \rho _i (T_e),   \label{fp2} 
\end{align}   

In addition to using Eq.~(\ref{eq:Entropdelta}), we add the $\sum_i^N W_i (T_e)$ term to increase the degree of freedom  of $S_{\text{emb}}(T_e)$.
As a result, $S_{\text{emb}}(T_e)$ can be expressed as
\begin{align}
    S_{\text{emb}}(T_e) & = \beta_1(T_e) \sum_i^N W_i (T_e) + \beta_2(T_e)  \sum_i^N W_i(T_e) ^2 \notag \\
                                 & = b_1(T_e) \sum_i^N  \sqrt{ \rho _i (T_e) } + b_2(T_e) \sum_i^N  \rho _i (T_e),   \label{eq:Entemb}
 \end{align}
 where Eq.~(\ref{eq:Emu2}) is used in the second equality, and $\beta_1(T_e), \beta_2(T_e), b_1(T_e)$ and $ b_2(T_e)$ are the fitting parameters.
 $S_{\text{tot}} (T_e)$ is then written as
\begin{align}                            
        S_{\text{tot}}(T_e)     & =  b_0 (T_e) +  b_1(T_e) \sum_i^N \sqrt{ \rho_i  (T_e)} +b_2 (T_e) \sum_i^N \rho _i (T_e).   \label{fp3}
\end{align}

A Dai potential~\cite{Dai_2006} is used to describe the function form of $V(r)$ and $\phi(r)$, which can be written as:
\begin{equation}
V(r) = \begin{cases}
                   \{ c_0(T_e) + c_1(T_e)r  \\
                   \hspace{0.5cm}   +c_2(T_e)r^2+c_3(T_e)r^3    &  \cdots \; r \le c(T_e)  \\
                   \hspace{0.5cm}   +c_4(T_e)r^4 \}      \{r-c(T_e) \}^2 \\
                    \hspace{2.5cm} 0   & \cdots \; r > c(T_e) ,   \label{eq:fp3}
         \end {cases}
\end{equation}
\begin{equation}
\phi(r) = \begin{cases}
                   \{r-d(T_e) \}^2+ d_0^2(T_e) \{r-d(T_e) \}^4 & \hspace{-0.4cm} \cdots r \le d(T_e) \\
                    \hspace{2cm} 0   & \hspace{-0.4cm}  \cdots  r > d(T_e) .   \label{eq:fp5}
         \end {cases}
\end{equation}
 
Fitting parameters:

 \begin{center}
 $ c(T_e),  d(T_e), a_0(T_e), a_1(T_e), a_2(T_e), b_0(T_e),  b_1(T_e), b_2(T_e),$
   $c_0(T_e), c_1(T_e), c_2(T_e),  c_3(T_e), c_4(T_e), d_0(T_e).$
 \end{center}
 
Here, $c(T_e)$ and $d(T_e)$ are the cutoff radii.
$a_0 (T_e)$ and $b_0 (T_e)$ represent $E_{\text{tot}}(T_e)$ and $S_{\text{tot}}(T_e)$ of isolated atoms, respectively.
These values are ideally the differences between the values of $E_{\text{tot}}(T_e)$ (or $S_{\text{tot}}(T_e)$) of isolated atoms at high $T_e$ and those of cold independent atoms.
Therefore, we assume that the excited atoms are emitted when non-thermal ablation occurs.
 This assumption is verified because the emission spectrum has been experimentally detected during the ablation process.~\cite{Harial_2013}

\subsection{Fitting methodology}
\begin{figure}[bp]
  \begin{center}
    \begin{tabular}{c}

      \begin{minipage}{1.0\hsize}
        \begin{center}
          \includegraphics[clip, width=7cm]{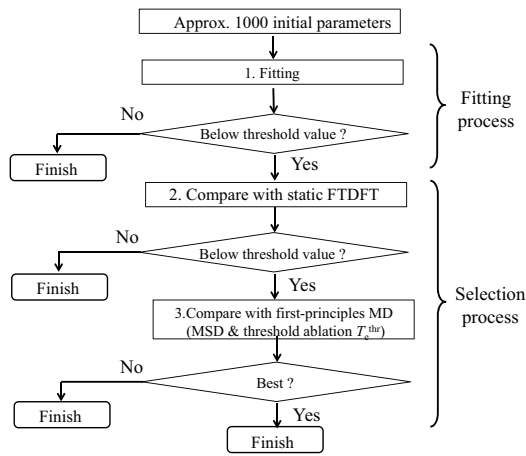}
          \hspace{1.6cm}
       
        \end{center}
      \end{minipage} \\
	
   \end{tabular}
    \caption{Flow chart to determine the $T_e$-dependent IAP.
              }
    \label{fig:fitting_flow}
  \end{center}
\end{figure}

Here we explain the flow used to determine the parameters of the $T_e$-dependent IAP for MD simulations of laser ablation.
In this study, the following two processes were performed (Figure~\ref{fig:fitting_flow}); the first is a fitting process and the other is a selection process.
In the fitting process, parameter fitting of $E_{\text{tot}}(T_e)$ and $F_{\text{tot}}(T_e)$ are carried out to select a candidate IAP.
In the selection process, the best parameters are selected from the candidates.  
Details are given next.

\subsubsection{Fitting process}
\label{sec:fitting_pro}

\begin{figure}[tbp]
  \begin{center}
    \begin{tabular}{c}

      \begin{minipage}{1.0\hsize}
        \begin{center}
          \includegraphics[clip, width=7cm]{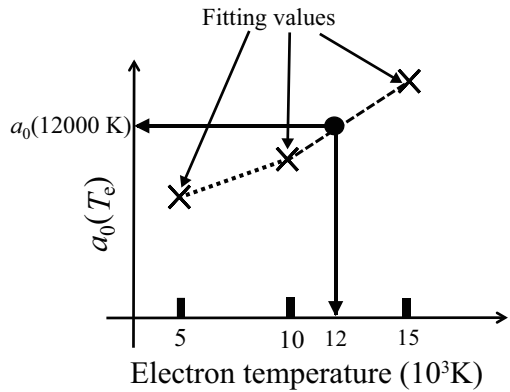}
       
        \end{center}
      \end{minipage} \\
	
   \end{tabular}
    \caption{Example of the interpolated value of $a_0(12000\,\text{K})$.
              }
    \label{fig:interpolate}
  \end{center}
\end{figure}

\begin{figure}[tbp]
  \begin{center}
    \begin{tabular}{c}

      \begin{minipage}{1.0\hsize}
        \begin{center}
          \includegraphics[clip, width=6cm]{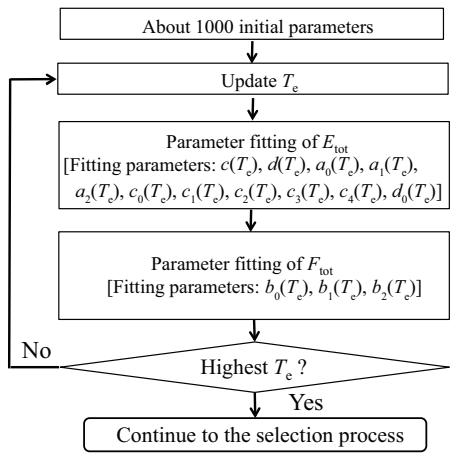}
       
        \end{center}
      \end{minipage} \\
	
   \end{tabular}
    \caption{Detail of ``1. Fitting''  in Fig.~\ref{fig:fitting_flow}.
              }
    \label{fig:fitting_flow2}
  \end{center}
\end{figure}

Here, we explain the details of the fitting process.
Parameter fitting of $E_{\text{tot}}(T_e)$ and $F_{\text{tot}}(T_e)$ was performed at $300\,\text{K}$, and from $T_e =5000$ to $50000\,\text{K}$ with an increment of $5000\,\text{K}$.
Parameter values of the other $T_e$ are estimated by linear interpolation. 
A simple example of the interpolation is represented in Fig.~\ref{fig:interpolate}.
To obtain appropriate $E_{\text{tot}}(T_e)$ and $F_{\text{tot}}(T_e)$ for the interpolation, the values of the fitting parameters should not be largely different from those near $T_e$.
Accordingly, the process in Fig.~\ref{fig:fitting_flow2} is followed to fit the parameters.
This process corresponds to ``1. Fitting'' in Fig.~\ref{fig:fitting_flow}.

First, parameter fitting was started from $T_e = 300\,\text{K}$.
We intend to obtain both appropriate $F_{\text{tot}}(T_e)$ and $E_{\text{tot}}(T_e)$; therefore, the parameter fitting of these values are carried out separately.
Parameter fitting of $E_{\text{tot}}(T_e)$ is performed at each $T_e$ before the fitting of $F_{\text{tot}}(T_e)$, i.e., the values of $c(T_e)$,  $d(T_e)$, $a_0(T_e)$, $a_1(T_e)$, $a_2(T_e)$, $c_0(T_e)$, $c_1(T_e)$, $c_2(T_e)$, $c_3(T_e)$, $c_4(T_e)$, and $d_0(T_e)$ are determined first. 
Next, parameter fitting of the other parameters [$b_0(T_e), b_1(T_e), b_2(T_e)$] is performed.
In the fitting process of $F_{\text{tot}}(T_e)$, the other parameters, such as $c(T_e)$ and $d(T_e)$, are fixed.
We then move on to parameter fitting of the next higher $T_e$.
The initial values of the parameters are then set to the values obtained through the previous parameter fitting.
In this way, parameter fitting is performed until the highest electron temperature $T_e=50000\,\text{K}$. 
These parameter fittings are performed approximately $1000$ times from the different initial values.

Fitting data [$E_{\text{tot}}(T_e)$ and $F_{\text{tot}}(T_e)$] are calculated using the VASP code~\cite{Kresse_1996,Kresse2_1996,Kresse_1993} based on the FTDFT. 
The number of fitting data $N_\text{fit}$ is $50$ at each $T_e$, and these data consist of calculation results of the fcc structures, small displacement structures from the fcc structures, and structures created by first-principles MD simulation.
The projector augmented wave (PAW)~\cite{Blochl_1994,Kresse_1999} method and the generalized gradient approximation (GGA) with the Perdew-Burke-Ernzerhof (PBE) exchange-correlation functional in the ground state ($T_e=0$) are used in the calculations.
Some studies have investigated the form of  the exchange-correlation energy ($ \Omega _{\text{xc}}$) and the effect of finite-temperature on it.~\cite{Dandrea_1986,Pittalis_2011,Eschrig_2010,Sjostrom_2014,Karasiev_2014,Karasiev_2016,Dornheim_2016,Burke_2016}
A theoretical study~\cite{Sjostrom_2014} investigated the contribution of the temperature dependence of $ \Omega _{\text{xc}}$ in homogeneous electron gas (HEG) of various densities.
The calculation results showed that, in the case of the electron density of condensed copper (Cu), the difference between the calculated free energy using finite-temperature local density approximation (LDA)~\cite{Karasiev_2014} and that using the ground-state LDA functional~\cite{Perdew_1981}  was less than $1\%$ at $T_e <50000\,\text{K}$.
The focus here is on laser-irradiated metals using relatively low fluence lasers, where $T_e$ is almost always $T_e < 50000\,\text{K}$; therefore, the finite-temperature effect of $ \Omega _{\text{xc}}$ is expected to be negligible.
Zero-temperature exchange-correlation functionals were thus used throughout our study.
This assumption is called the ground-state approximation, and has been widely employed to investigate phenomena caused by irradiation with an ultrashort pulse laser.
Calculation results based on this assumption have successfully reproduced experimental results, such as the disappearance of Jahn-Teller distortion,~\cite{Fritz_2007} ultrafast melting,~\cite{Daraszewicz_2013} and bond hardening.~\cite{Ernstorfer_2009}

The cohesive energy of the PBE calculation is $3.68\,\text{eV}$, which is slightly larger than the experimental value ($3.49\,\text{eV}$~\cite{Kittle,Kaxiras}).
The value of PBE is better than the values calculated using PBEsol  ($4.26\,\text{eV}$) and LDA ($4.67\,\text{eV}$).
The conserved energy is set to $1.0\times10^{-4}\,\text{eV}$, and the occupation number of the highest energy band is less than $0.001$. 
To adjust the value of the energy of the isolated atom limit $E_\text{atom}$ at low $T_e$ to $0$, $E_\text{atom}$ is subtracted from all values of $E(T_e)$ and  $F(T_e)$.
The electronic structure calculations were performed with a cutoff energy of $480\,\text{eV}$ for the plane-wave basis and Brillouin-zone $k$-points sampling of a $8\times 8\times 8$ Monkhorst-Pack mesh for the fcc structures and the small displacement structures.
Atom dynamics simulation was also performed at high $T_e$ for a small system with $108$ atoms.
To represent a thin film (ca. $10\,\text{nm}$), the slab model was employed, and the lattice constants of the computational cell were fixed to $x=10.845\,\AA$, $y=10.845\,\AA$, and $z=36.15\,\AA$. 
The periodic boundary conditions were applied in all directions.
The time step was $3\,\text{fs}$ and $k$-points sampling was $4\times 4\times1$.
Prior to simulations at high $T_e$,  atoms were thermalized using the Nos$\acute{\text{e}}$-Hoover thermostat~\cite{Hoover_1985} at $300\,\text{K}$ for more than $3\,\text{ps}$.

The root mean square error (RMSE) was used as the evaluation function for the parameter fitting.
The definition of the RMSE is
\begin{eqnarray}
   \text{RMSE} = \frac{1}{N_{\text{fit}}} \sum_q^{N_\text{fit}}  \sqrt{ \left( \frac{E_q^{\text{FTDFT}}}{N_q^\text{atom}} \right)^2 - \left( \frac{E_q^{\text{IAP}}}{N_q^\text{atom}} \right) ^2 }, \label{eq:RMSE}
\end{eqnarray}
 where $E^{\text{FTDFT}}$ is the internal energy or the free energy of the FTDFT calculations and $E^{\text{IAP}}$ is that of the IAP calculations.
 $q$ is the index of the structures and $N^\text{atom}_q$ is the number of atoms in the $q$-th structure. The non-linear mean square method was used for the parameter fitting.
The Gauss-Seidel method was used to solve the non-linear equations.

\subsubsection{Selection process}
\label{sec:selection}

 In the fitting process, many IAP [$E_{\text{tot}}(T_e)$ and $F_{\text{tot}}(T_e)$] were obtained as candidates.
 In this process, the best IAP was selected from these IAP for the ablation simulation.
 
The best IAP was considered as the potential that could reproduce the FTDFT results of the cohesive energy, the lattice constant, the bulk modulus, the phonon dispersion, the means square displacement (MSD), and the ablation threshold electron temperature $T_e^\text{thr}$ because these values are expected to be important for a valid ablation simulation.
The definition of the MSD is
  \begin{eqnarray}
\text{MSD} =  \frac{1}{N^{\text{atom}}_q} \sum_i^{N^{\text{atom}}_q} (\bm{r}_i(t) - \bm{r}_i(0))^2, \label{eq:MSD}
\end{eqnarray}
where $t$ is the elapsed time after $T_e$ is increased, and $\bm{r}_i(t)$ represents the position of the $i$-th atom at $t$.

The cohesive energy, the lattice, and the bulk modulus are standard physical properties that should be reproduced by any IAP.
The phonon dispersion represents the stability of the structure and is related to the melting temperature, which is important to describe the spallation process.
The MSD is related with the diffusion of atoms and the expansion rate, and thus changes significantly with each phase, i.e., solid, liquid, and gas.
The importance of $T_e^\text{thr}$ in an investigation of the mechanism of ablation is obvious.

The cohesive energy, the lattice constant, and the bulk modulus $B$, are calculated by fitting the energy-volume ($EV$) curve to the following Murnaghan equation of state,~\cite{Poirier}
 \begin{equation}
	E(V) = E(V_0) \frac{BV}{B'(B' - 1)} \Biggl[  B'  \Big( 1- \frac{V_0}{V} \Big)+ \Big( \frac{V_0}{V} \Big)^{B'} - 1 \Biggr],    \label{eq:BM}
\end{equation}
where $V$, $V_0$, and $B'$ are the volume, equilibrium volume, and the derivative of $B$ with respect to pressure, respectively.
In this study, if all the errors of these values are below $20\%$, then the IAP is considered to be an appropriate potential.

Calculation conditions for the phonon dispersion are as follows. 
 Forces that act on atoms were calculated using the VASP code~\cite{Kresse_1996,Kresse2_1996,Kresse_1993} and phonon calculations were conducted using the ALAMODE package.~\cite{Tadano,Tadano_2014}
 The force constant was calculated using the frozen phonon method.
 The calculated cell is a $3\times3\times3$ supercell of the conventional unit cell of the fcc structure and the $k$-points sampling is $3\times3\times3$.
 We consider that if an imaginary phonon does not exist and the error of the maximum frequency is less than $150\,\text{cm}^{-1}$, then the IAP is regarded as an appropriate potential.
 It should be noted that this criterion is only applied to the phonon dispersion below $T_e^\text{thr}$ because the importance of the phonon dispersion above $T_e^\text{thr}$ is expected to be low.

First-principles MD simulations were performed to calculate the MSD and $T_e^\text{thr}$.
The details of the MD simulation are the same as those explained for the fitting process.
MD simulations were performed three times with different initial configurations.
When the bottom and surface atoms are within the cutoff radius of the IAP, ablation is considered to have occurred at this $T_e$. 
We consider that if the error of $T_e^\text{thr}$ is less than $2500\,\text{K}$, then this IAP is appropriate.
The MSD calculation results are used to determine the best IAP from all IAP that satisfy all the criteria explained above.

 \section{Results and Discussion}
\label{sec:result}

Here, we analyze the validity of the IAP obtained by comparison of the IAP calculation results with the FTDFT calculation results. 
 
 \subsection{Fitting accuracy}

\begin{figure}[tp]
  \begin{center}
    \begin{tabular}{c}

      \begin{minipage}{1.0\hsize}
        \begin{center}
          \includegraphics[clip, width=7cm]{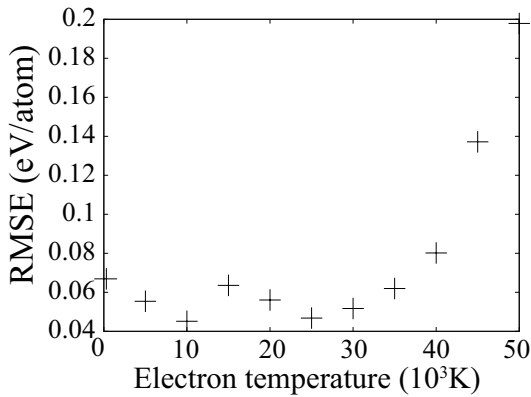}
       
        \end{center}
      \end{minipage} \\
	
   \end{tabular}
    \caption{ RMSE of the fitting results with the best IAP.
           Values of  RMSE are per atom. 
              }
    \label{fig:RMSE}
  \end{center}
\end{figure}

The RMSE of the fitting results with the best IAP is shown in Fig.~\ref{fig:RMSE}.
 Figure~\ref{fig:RMSE} shows that at low $T_e$, the IAP can reproduce the FTDFT results with an error of several $10\,\text{meV}$.
 We consider that the large error at high $T_e$ is due to occupation of the high energy 4$p$ states, which are unoccupied at low $T_e$.
 The developed functional form [Eq.~(\ref{fp2})] uses a rectangular model; therefore, it would be difficult to express this effect.

 \begin{figure*}[bth]
  \begin{center}
    \begin{tabular}{cccc}

      \begin{minipage}{0.25\hsize}
        \begin{center}
          \includegraphics[clip, width=4.5cm]{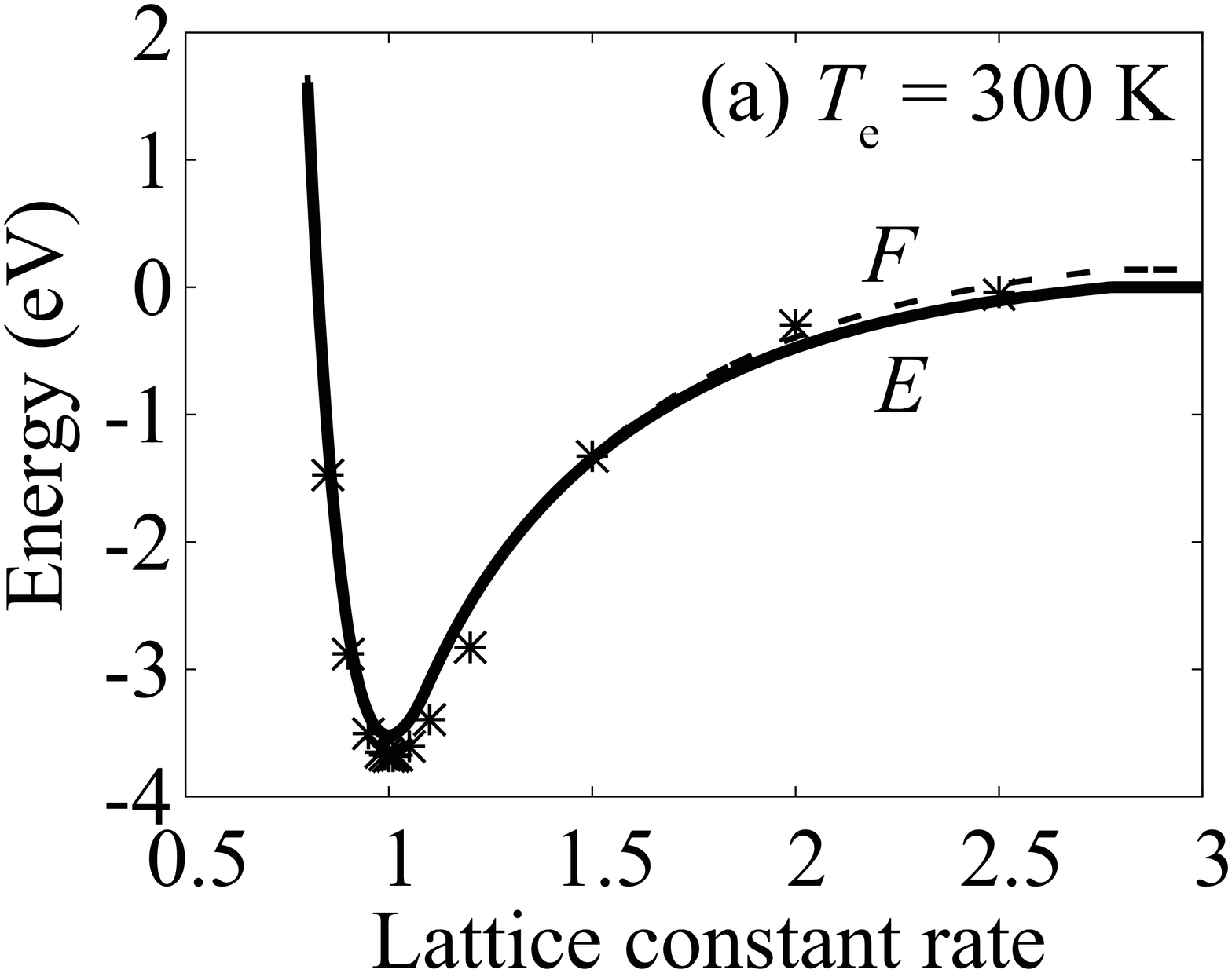}
        \end{center}
      \end{minipage}
      \begin{minipage}{0.25\hsize}
        \begin{center}
          \includegraphics[clip, width=4.5cm]{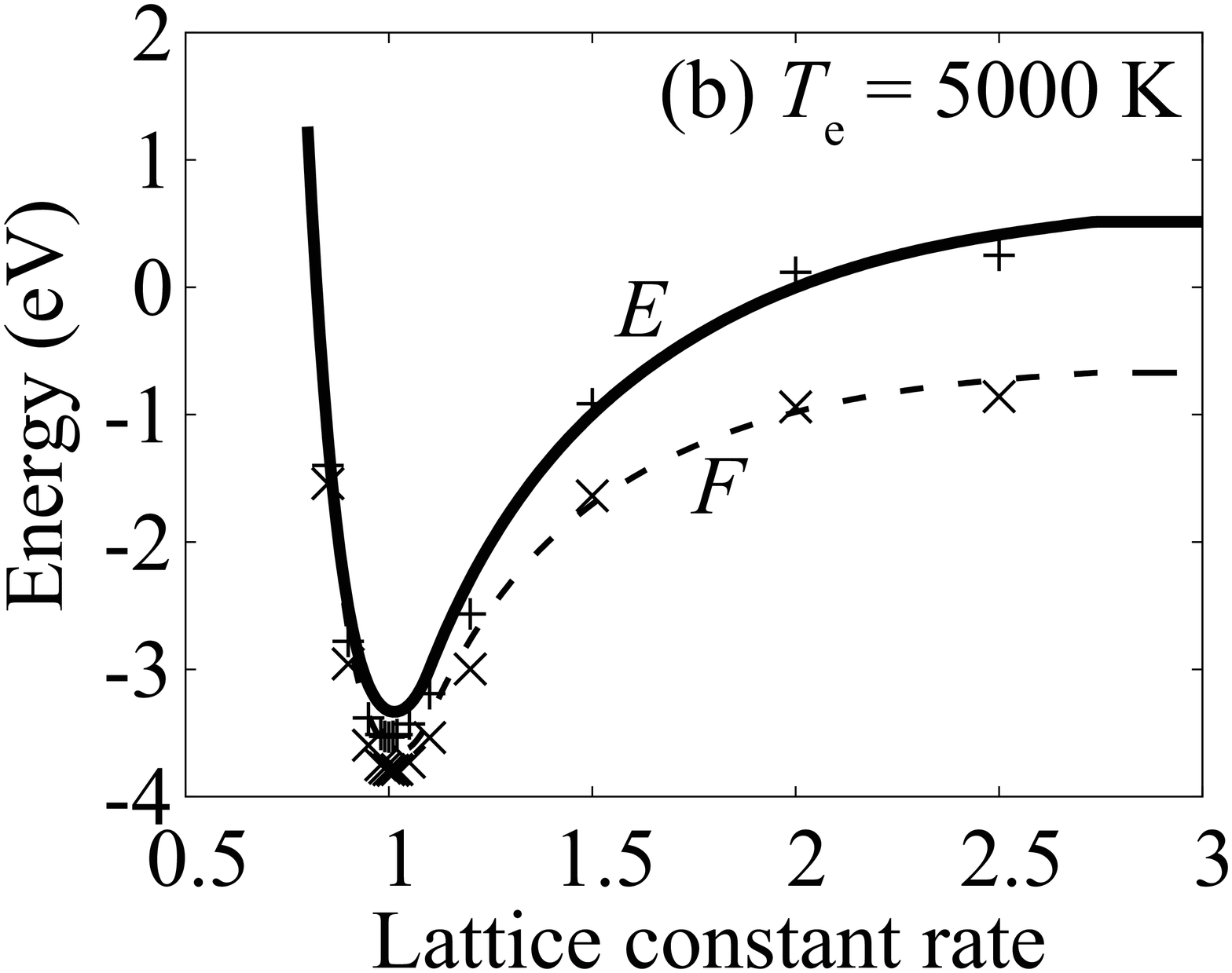}
        \end{center}  
      \end{minipage}
      \begin{minipage}{0.25\hsize}
        \begin{center}
         \includegraphics[clip, width=4.5cm]{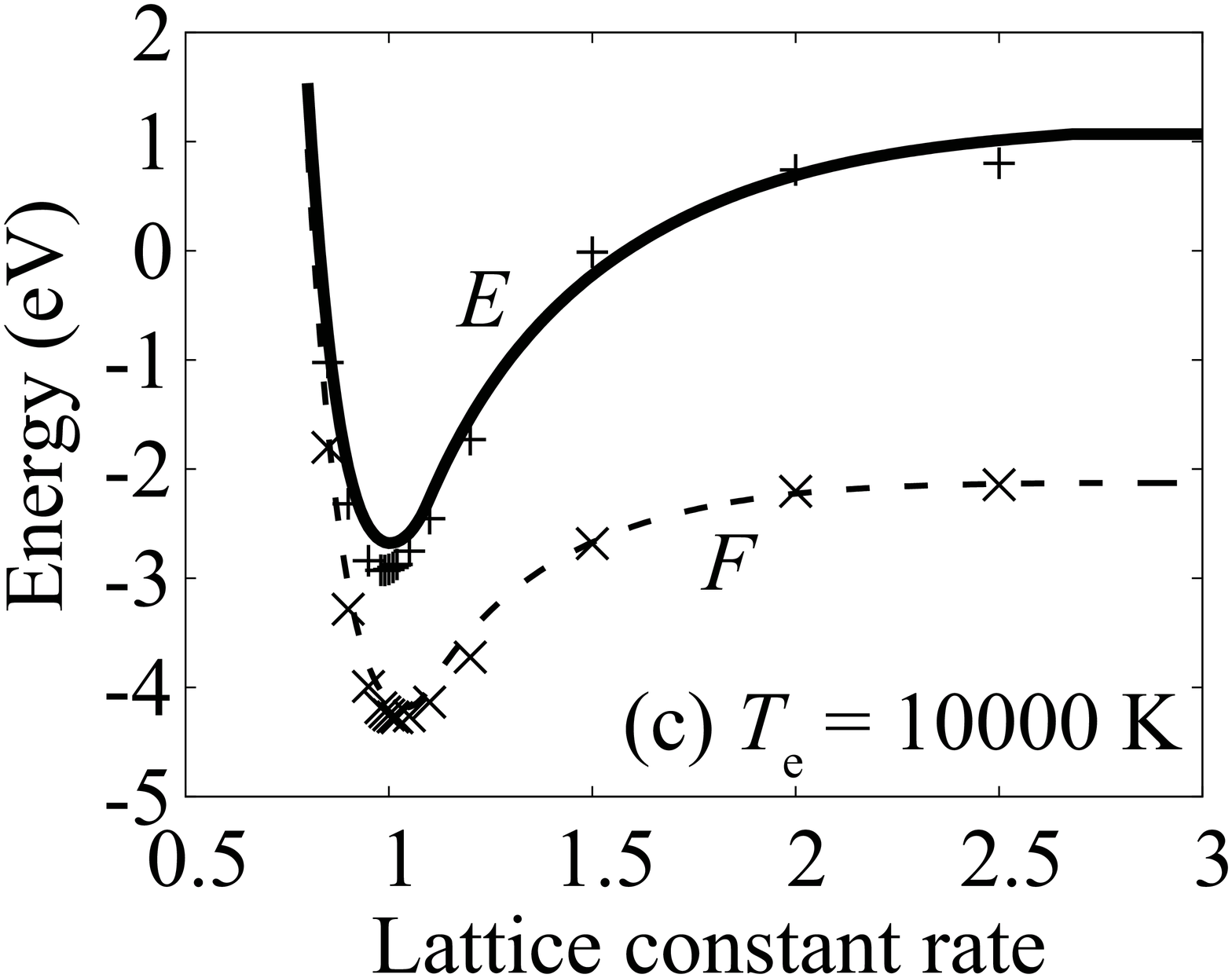}
        \end{center}
      \end{minipage}
      \begin{minipage}{0.25\hsize}
        \begin{center}
          \includegraphics[clip, width=4.5cm]{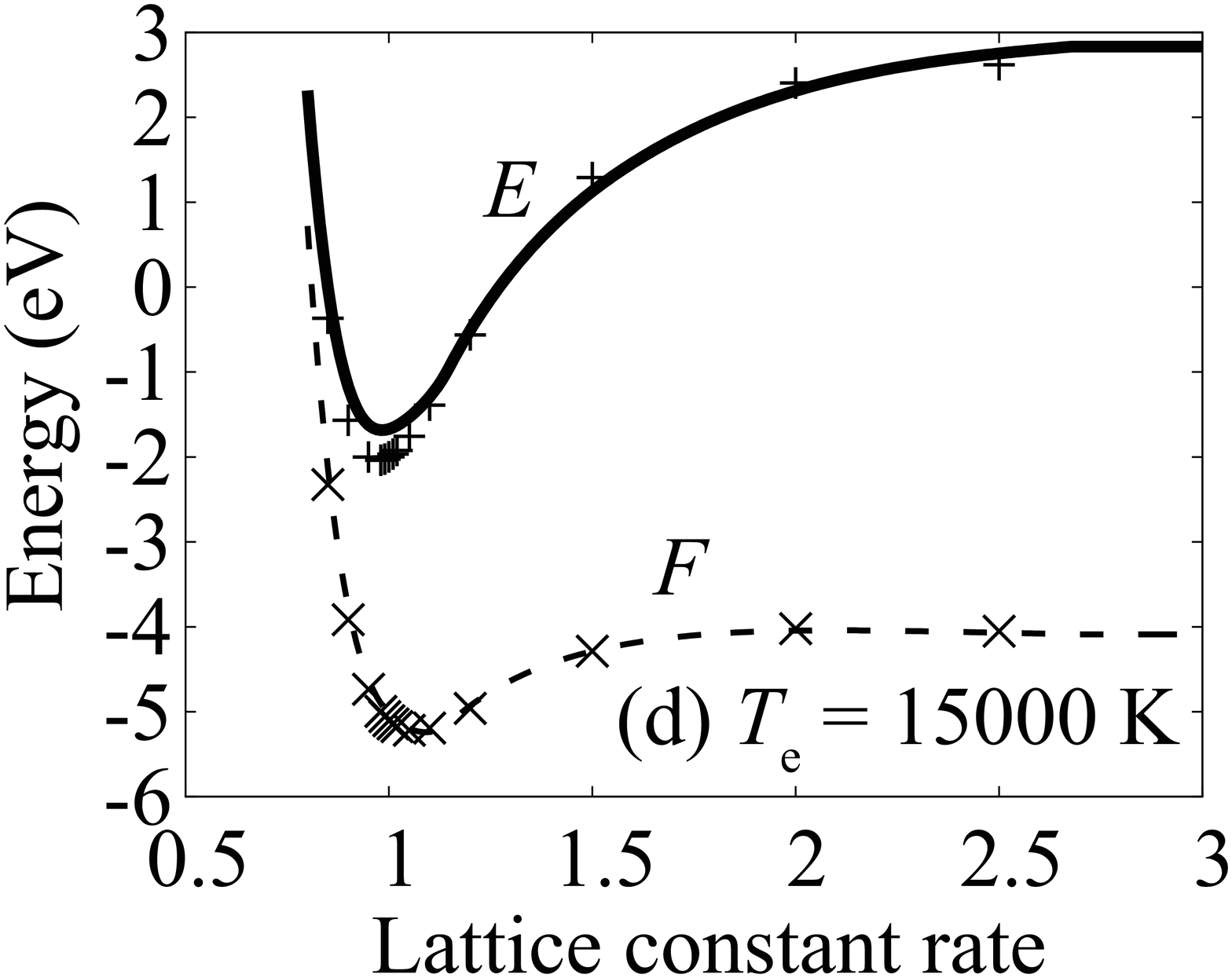}
        \end{center}
      \end{minipage}
      \\
      \begin{minipage}{0.25\hsize}
        \begin{center}
          \includegraphics[clip, width=4.5cm]{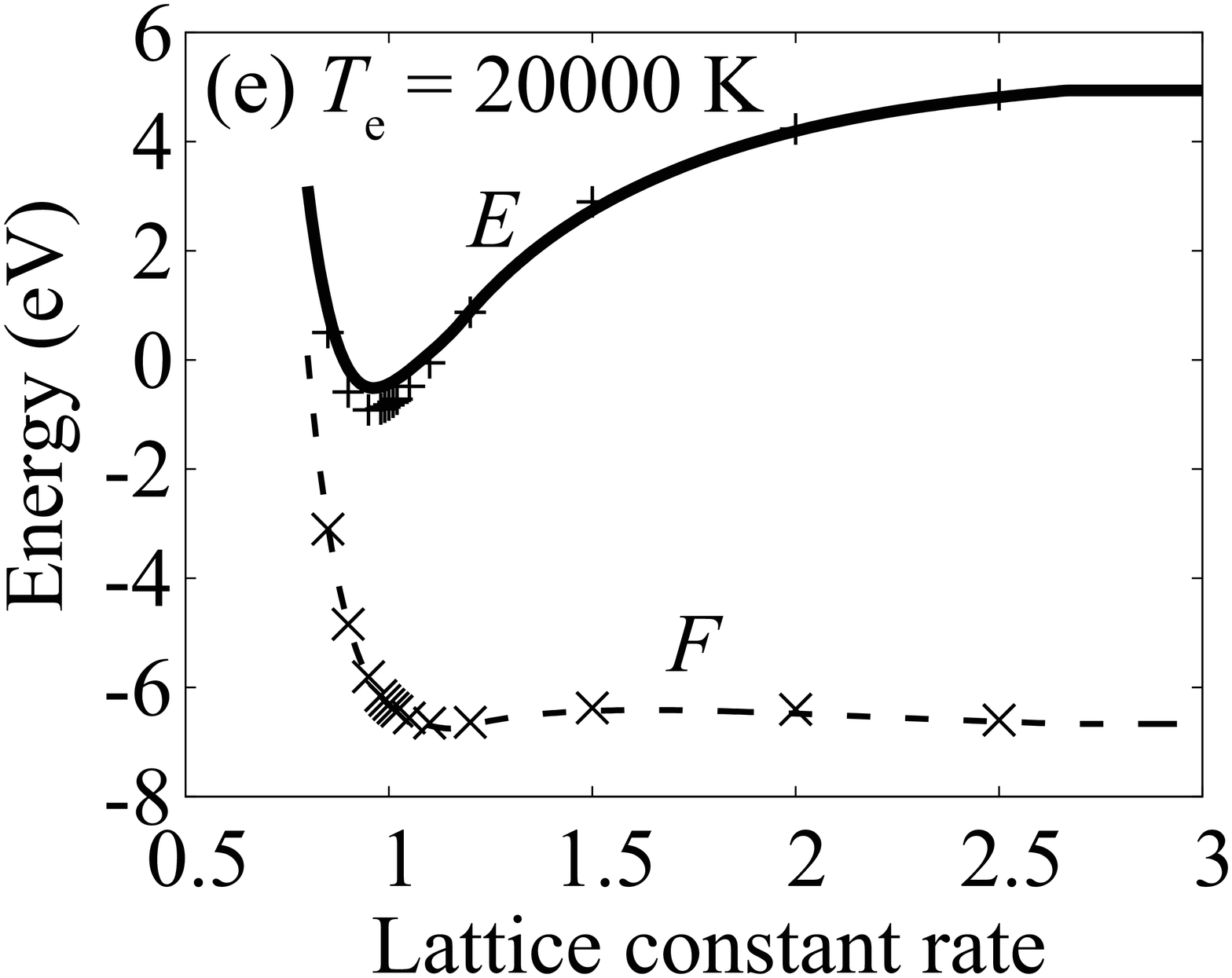}
        \end{center}
      \end{minipage}
      \begin{minipage}{0.25\hsize}
        \begin{center}
        \includegraphics[clip, width=4.5cm]{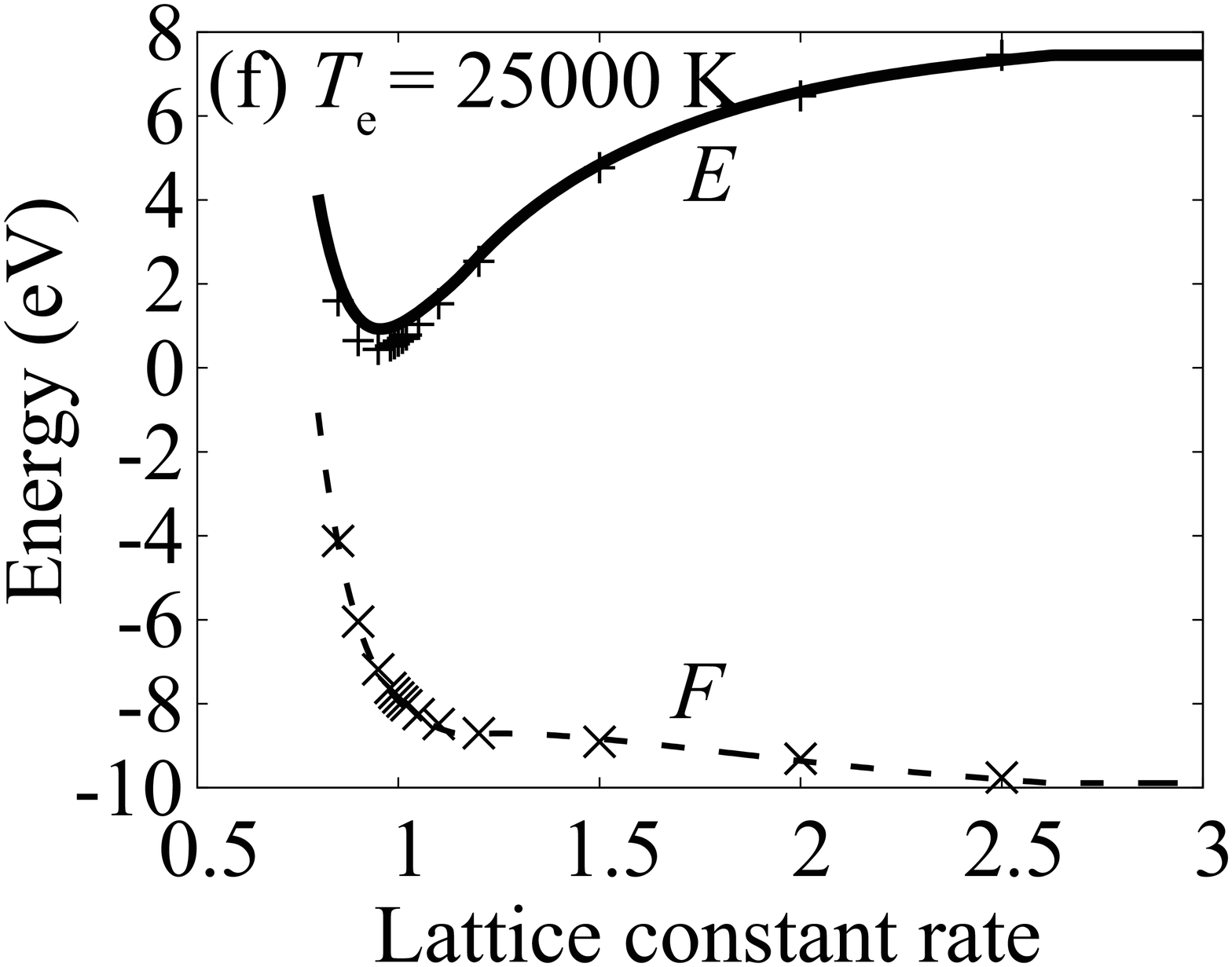}
        \end{center}
      \end{minipage}
      \begin{minipage}{0.25\hsize}
        \begin{center}
          \includegraphics[clip, width=4.5cm]{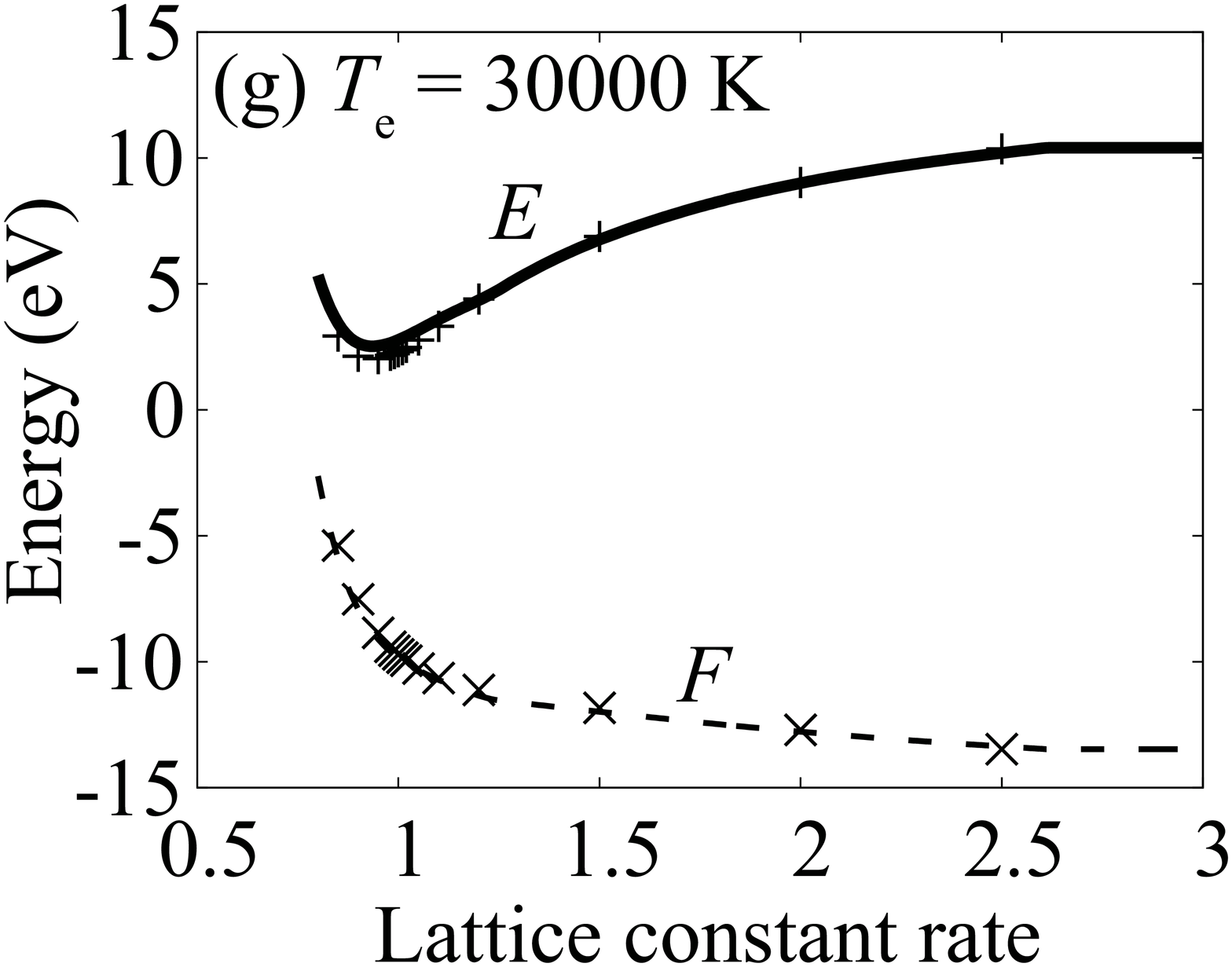}
        \end{center}
      \end{minipage}
      \begin{minipage}{0.25\hsize}
        \begin{center}
        \includegraphics[clip, width=4.5cm]{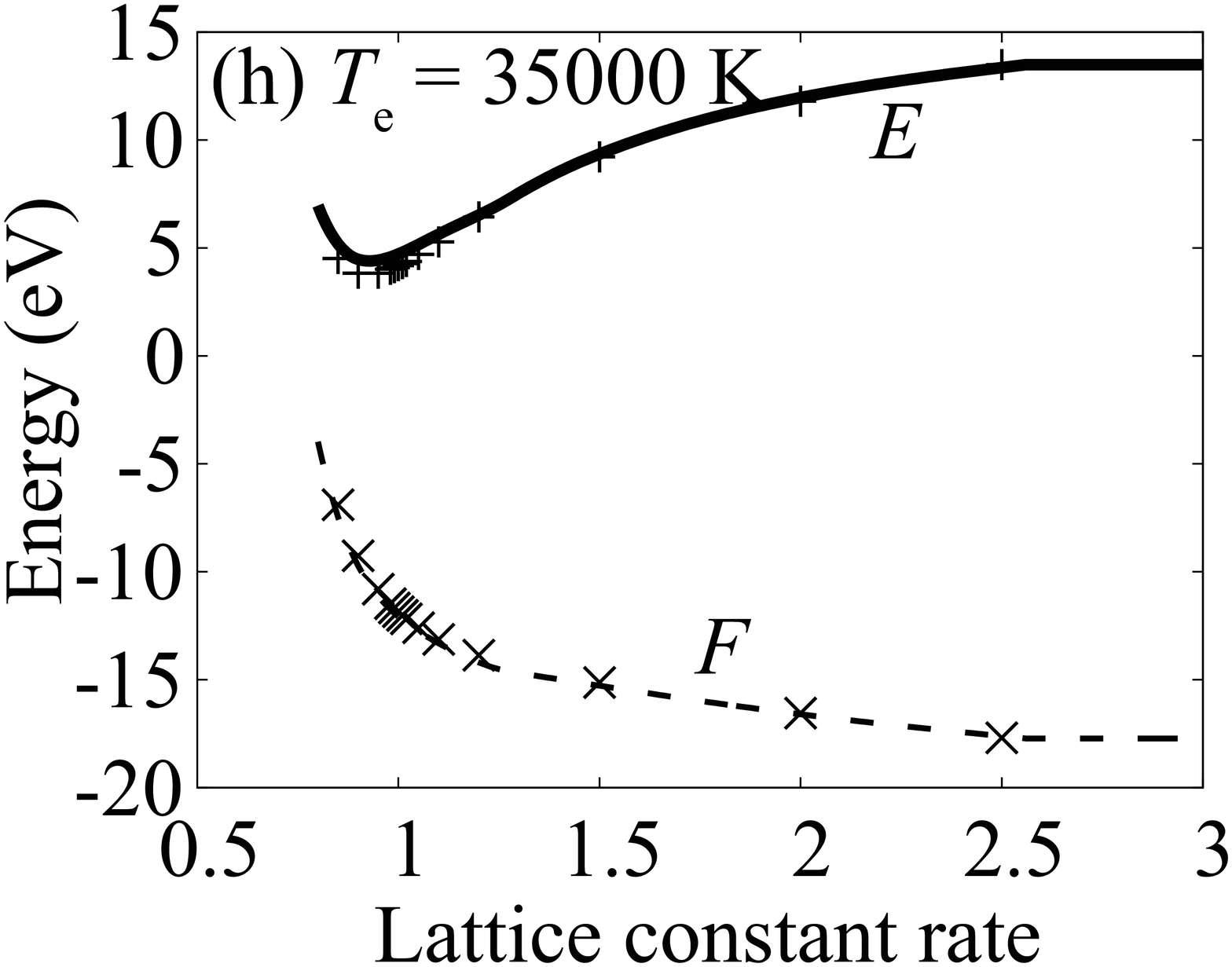}
        \end{center}
      \end{minipage}
      \\
      \begin{minipage}{0.25\hsize}
        \begin{center}
          \includegraphics[clip, width=4.5cm]{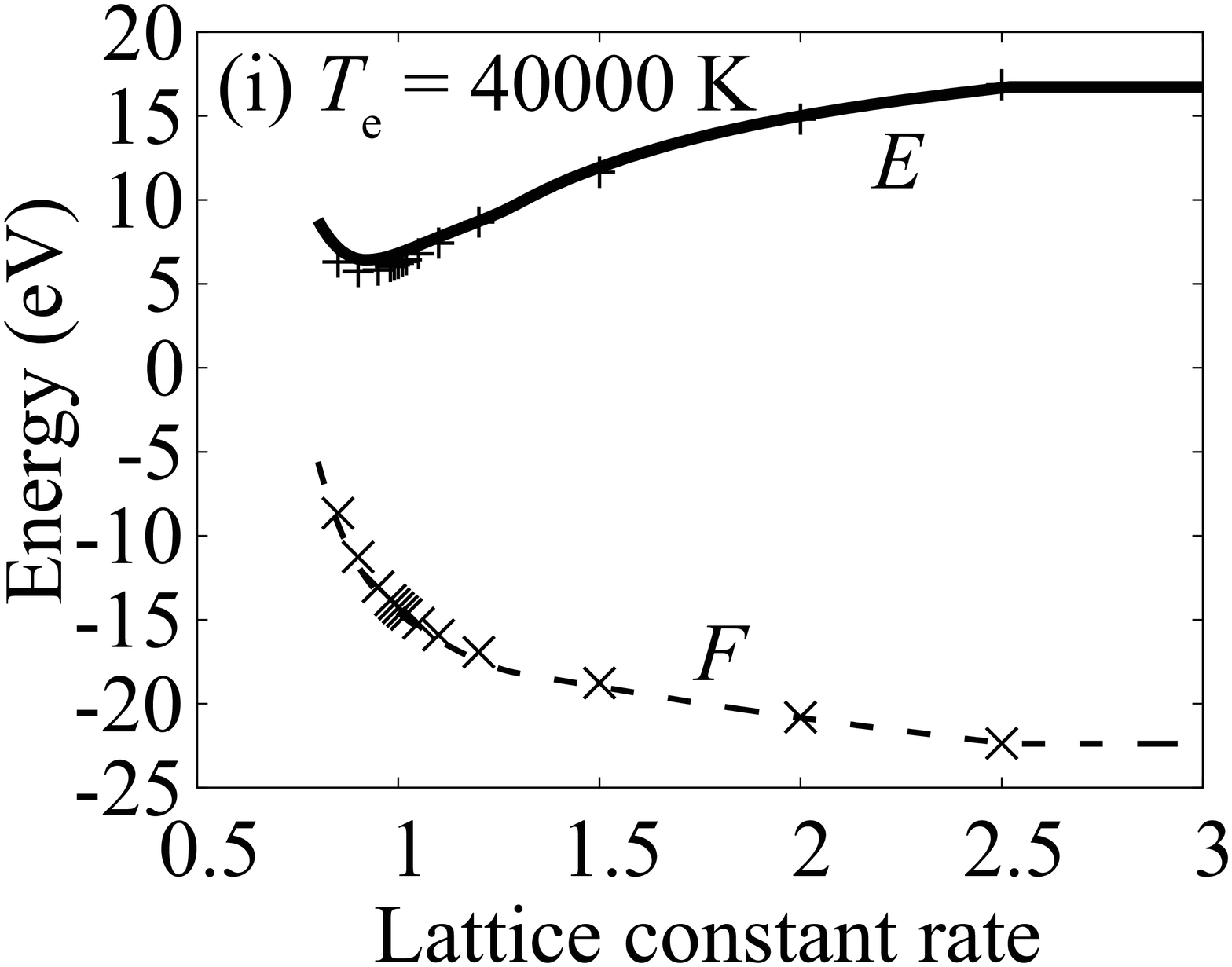}
        \end{center}
      \end{minipage}
      \begin{minipage}{0.25\hsize}
        \begin{center}
        \includegraphics[clip, width=4.5cm]{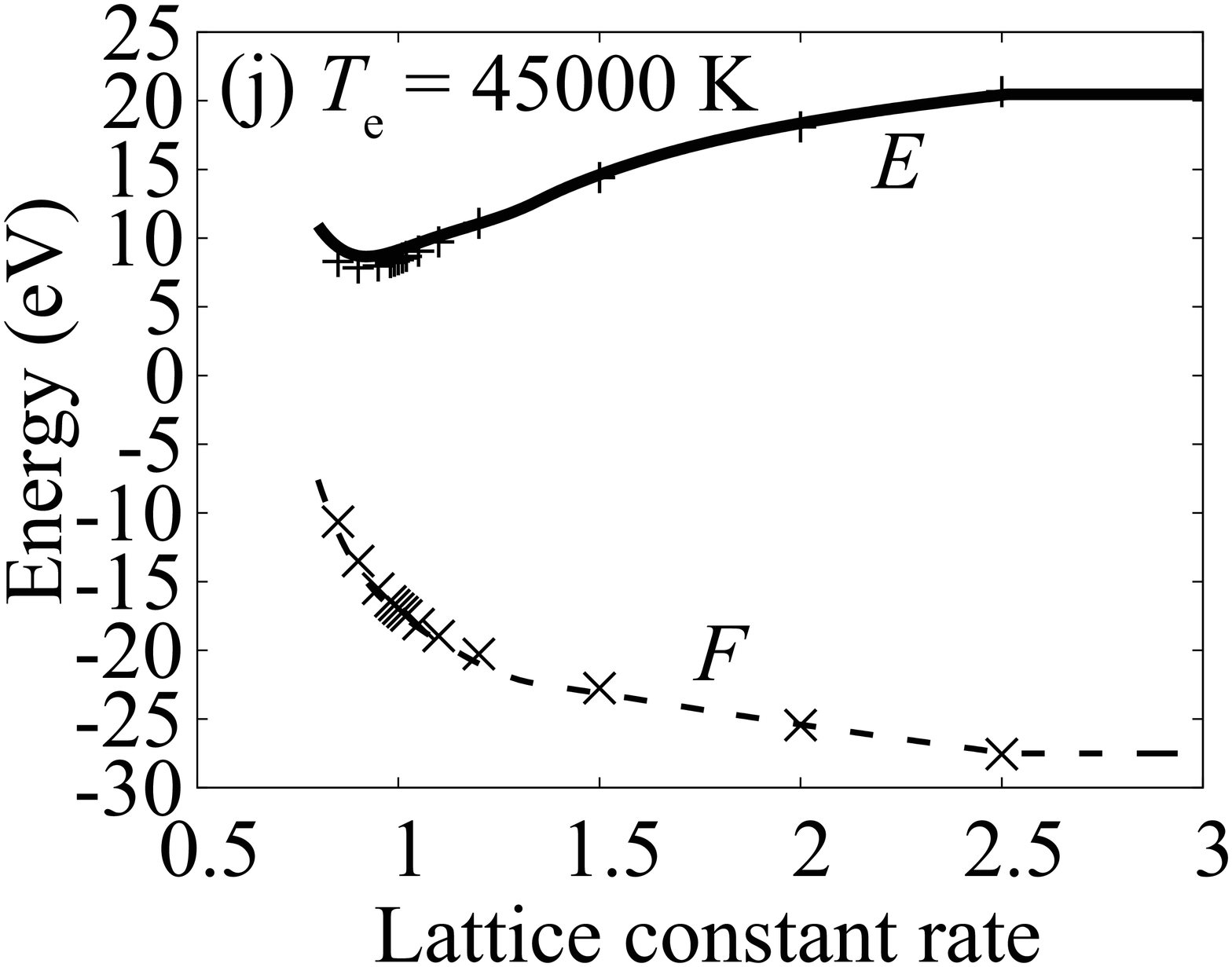}
        \end{center}
      \end{minipage}
      \begin{minipage}{0.25\hsize}
        \begin{center}
          \includegraphics[clip, width=4.5cm]{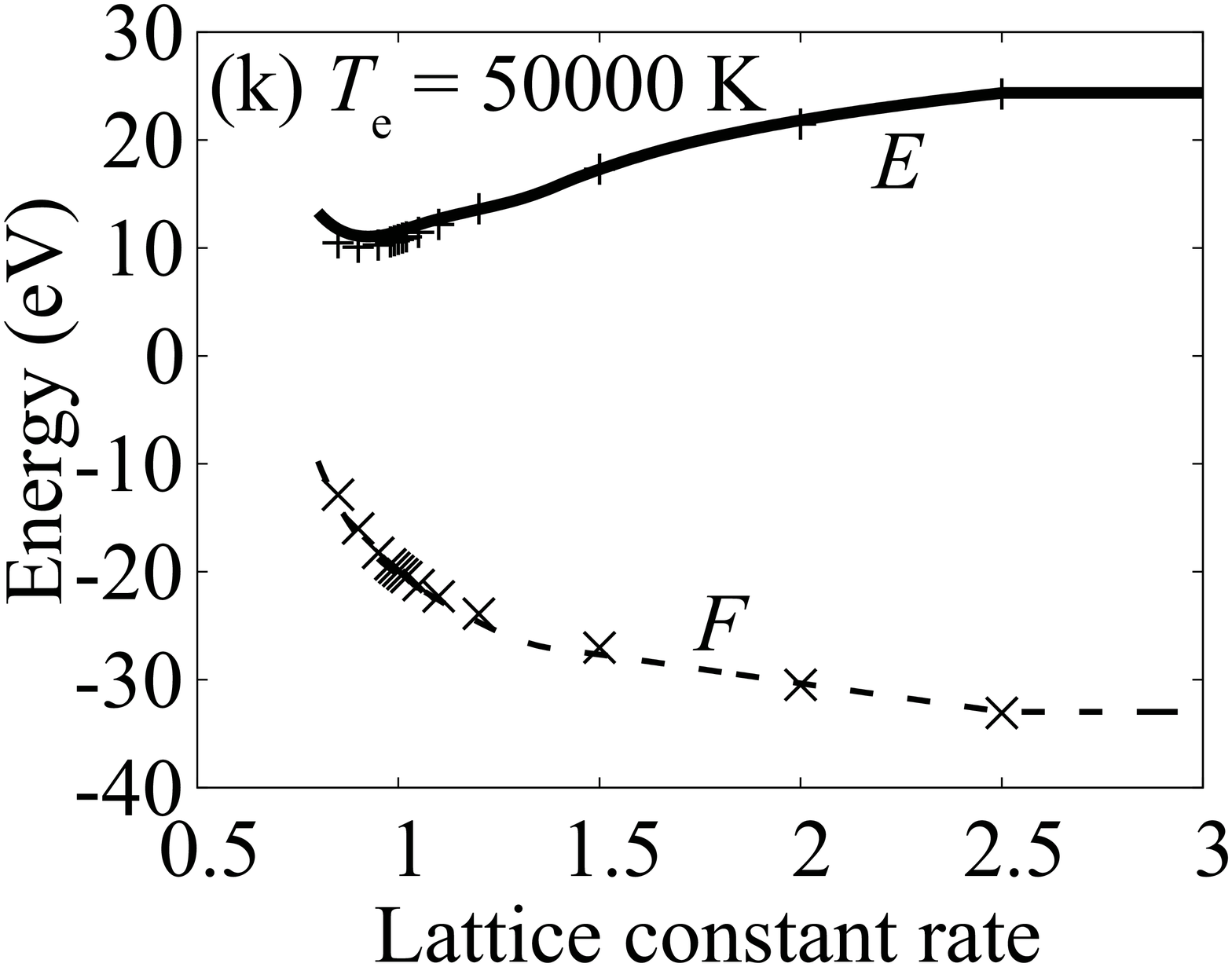}
        \end{center}
      \end{minipage}
      \begin{minipage}{0.25\hsize}
        \begin{center}
        \includegraphics[clip, width=2cm]{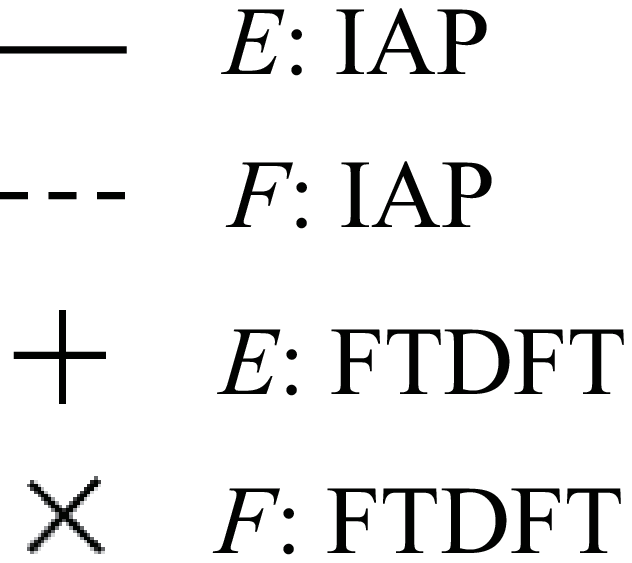}%
        \end{center}
      \end{minipage}
   \end{tabular}
      \caption{Volume dependence of $E$ and $F$.
      Each horizontal axis indicates the rate of the lattice constant with respect to the equilibrium lattice constant at $T_e=300\,\text{K}$.
      Plus and cross marks represent the FTDFT results for $E$ and $F$, respectively.
      Solid and dashed lines represent IAP calculation results for $E$ and $F$, respectively.
     }
    \label{fig:EFSV}
  \end{center}
\end{figure*}

\begin{table}[bp]
\begin{center}
\caption{Values of lattice constant, cohesive energy, and the bulk modulus at $T_e=300\,\text{K}$.}
    \small
      \scalebox{1}[1]{
\begin{tabular}{cccccccccc} \hline 
                              & IAP & FTDFT & Exp. \\ \hline\hline
Lattice constant ($\AA$)   & $3.613$  & $3.634$ & $3.615$~\cite{Person}  \\ 
Cohesive energy (eV)   & $3.52$  & $3.68$ & $3.49$~\cite{Kittle}  \\ 
Bulk modulus    (GPa)    & $164.6$ & $137.6$  & $142$~\cite{Simmons}  \\ \hline
\end{tabular}
}

\label{tb:lcb}
\end{center}
\end{table}

\subsection{Volume dependence of $E$ and $F$ }
 
 Figure~\ref{fig:EFSV} shows the volume dependence of $E$ and $F$.
 The horizontal axis indicates the rate of the lattice constant with respect to the equilibrium lattice constant at $T_e=300\,\text{K}$.
 Figure~\ref{fig:EFSV} shows that the IAP calculation results agree well with the FTDFT calculation results.
 
 The results for the cohesive energy, lattice constant, and bulk modulus at $T_e=300\,\text{K}$ are summarized in Table~\ref{tb:lcb}.
 The result of the IAP calculations overestimates the value of the bulk modulus, whereas the values of the lattice constant and the cohesive energy are very close to the experimental values.

 \subsection{Phonon dispersion}
 
Figure~\ref{fig:phonon} shows the $T_e$-dependent phonon dispersion by FTDFT calculations and IAP calculations.
The IAP calculation results in an overestimation of the phonon frequency by  ca. $ 30 \, \%$.
This discrepancy can be expected from the calculation results of the bulk modulus of IAP, which is overestimated by the FTDFT calculation (Table~\ref{tb:lcb}).
In addition, the $T_e$ dependence behavior with respect to the phonon hardness is different between these results.
The cause of this discrepancy can be considered to come from the small energy scale of the phonon with respect to the cohesive energy.
As a result, the MD simulations using the IAP at relatively low $T_e$ cannot reproduce the FTDFT simulation.
However, this discrepancy has little effect on the simulations of laser ablation because the ablation $T_e$, which will be explained below (Sec.~\ref{sec:Tethr}), is more important than this value.
In addition, the imaginary phonon does not exist below 15000$\,$K so that the lattice stability, which is important for the spallation process, is verified below 15000$\,$K.
The IAP reproduces the result of the FTDFT calculation in that the phonon instability does not occur, even at the electron temperature around which ablation occurs.
Therefore, it can be expected that, similar to simulations based on FTDFT, simulations using IAP do not produce structural change below the ablation threshold temperature.
 To calculate the phonon dispersion more accurately, use of a neural network potential would be appropriate, which has been reported to result in accurate predictions with finite temperature energy calculations of FTDFT.~\cite{Zhang_2020}

\begin{figure}[btp]
    \begin{tabular}{cc}

      \begin{minipage}{0.5\hsize}
        \begin{center}
          \includegraphics[clip, width=4.0cm]{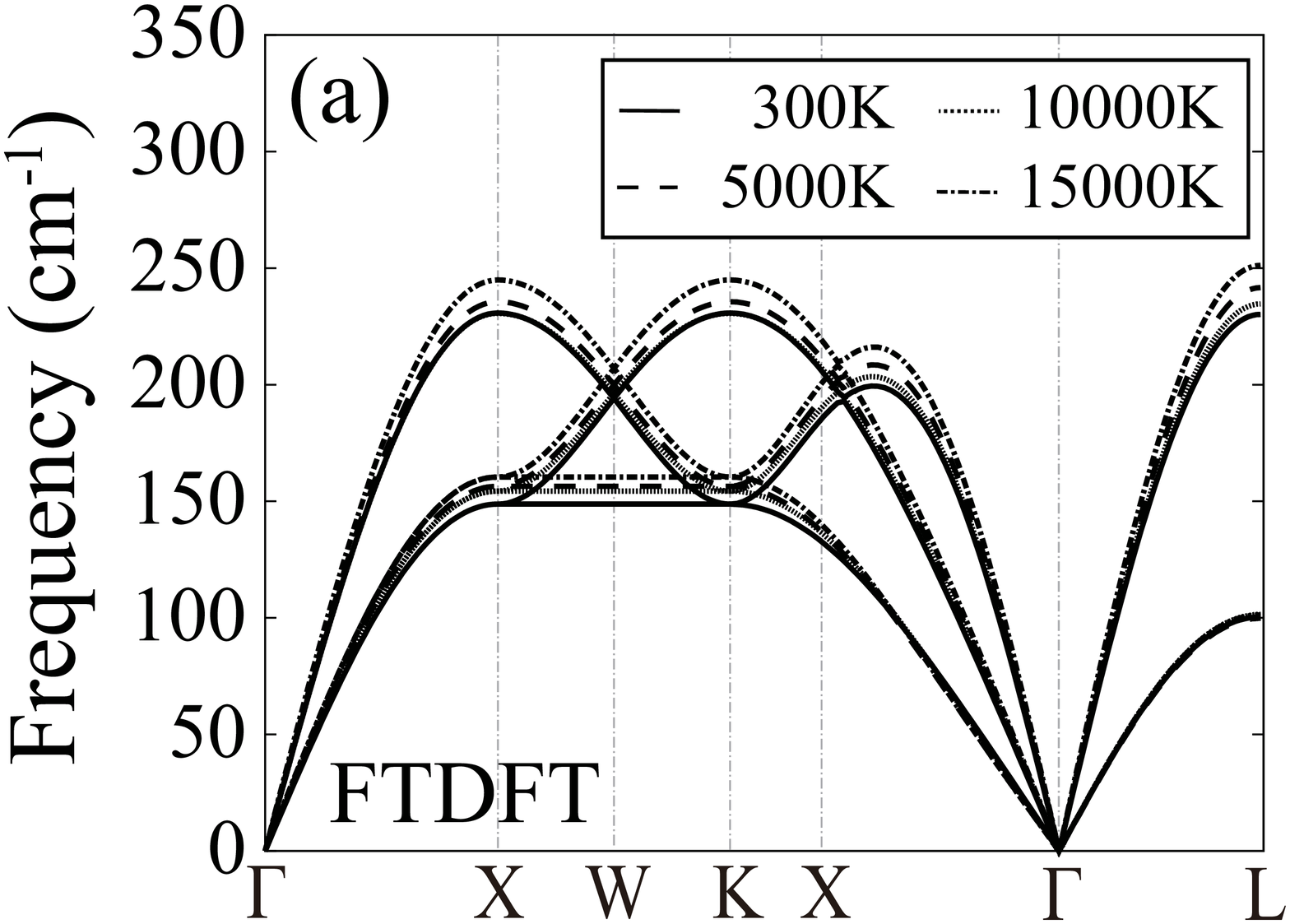}
        \end{center}
     \end{minipage} 

      \begin{minipage}{0.5\hsize}
        \begin{center}
          \includegraphics[clip, width=4.0cm]{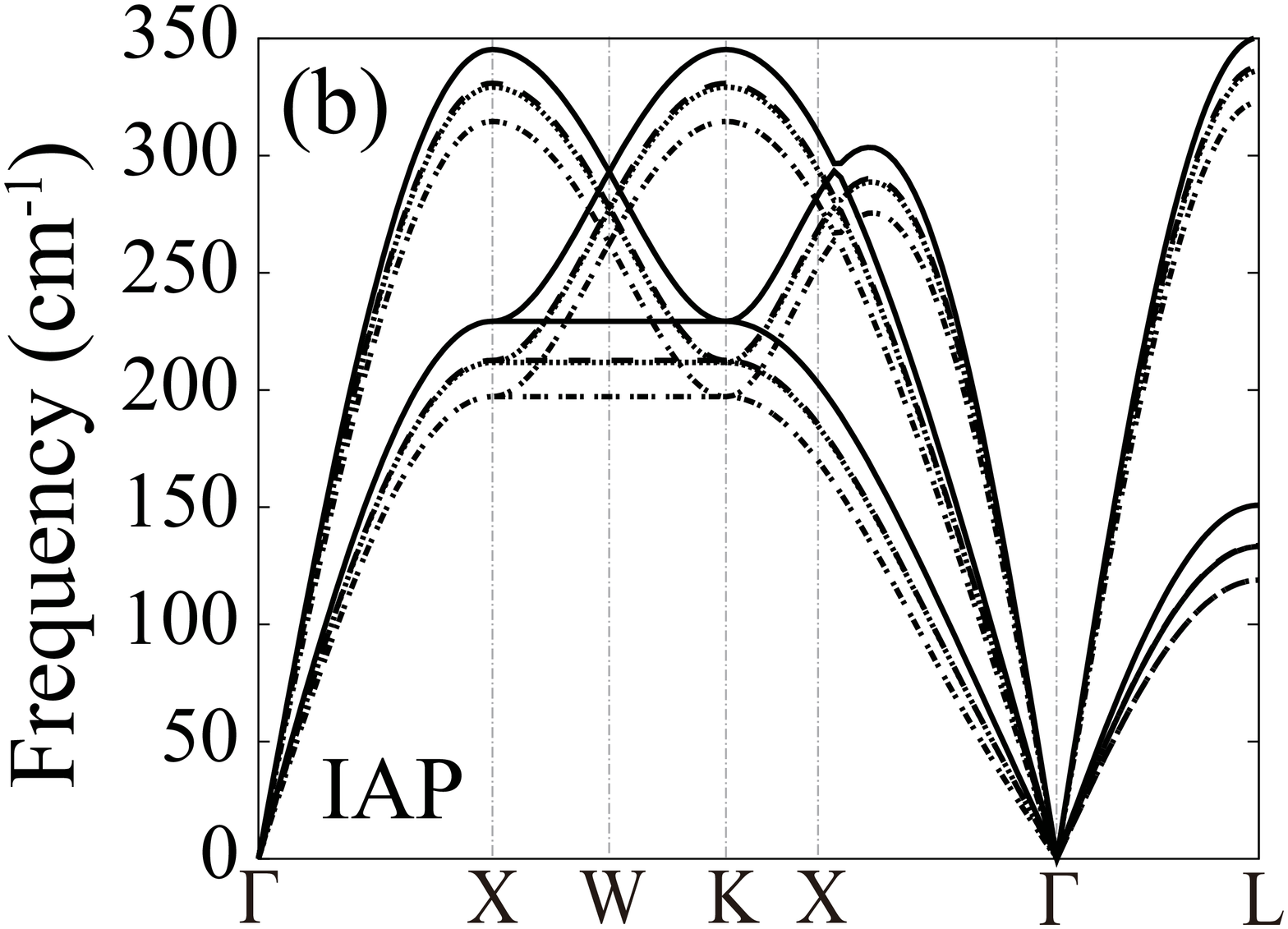}
             \end{center}
     \end{minipage} 
   \end{tabular}

    \label{fig:phonon}
         \caption{
    Calculation results of phonon dispersion from (a) FTDFT calculations and (b) IAP calculations.
    Solid, dashed, dotted, and chain lines represent calculation results at $T_e=300$, $5000$, $10000$, and $15000\,\text{K}$, respectively.
    Each high symmetry point represents $\Gamma= [0,0,0]$, $\text{X}=[1/2,0,1/2]$, $\text{W}=[1/2,1/4,3/4]$, $\text{K}=[1/2, 1/2, 1]$, and $\text{L}=[1/2,1/2,1/2]$.
              }
       \label{fig:phonon}
\end{figure}

 \subsection{Mean square displacement (MSD)}
\label{sec:MSD}
  
\begin{figure}[tbp]
\begin{minipage}{1.0\hsize}
    \begin{tabular}{cc}

      \begin{minipage}{0.5\hsize}
        \begin{center}
          \includegraphics[clip, width=4.0cm]{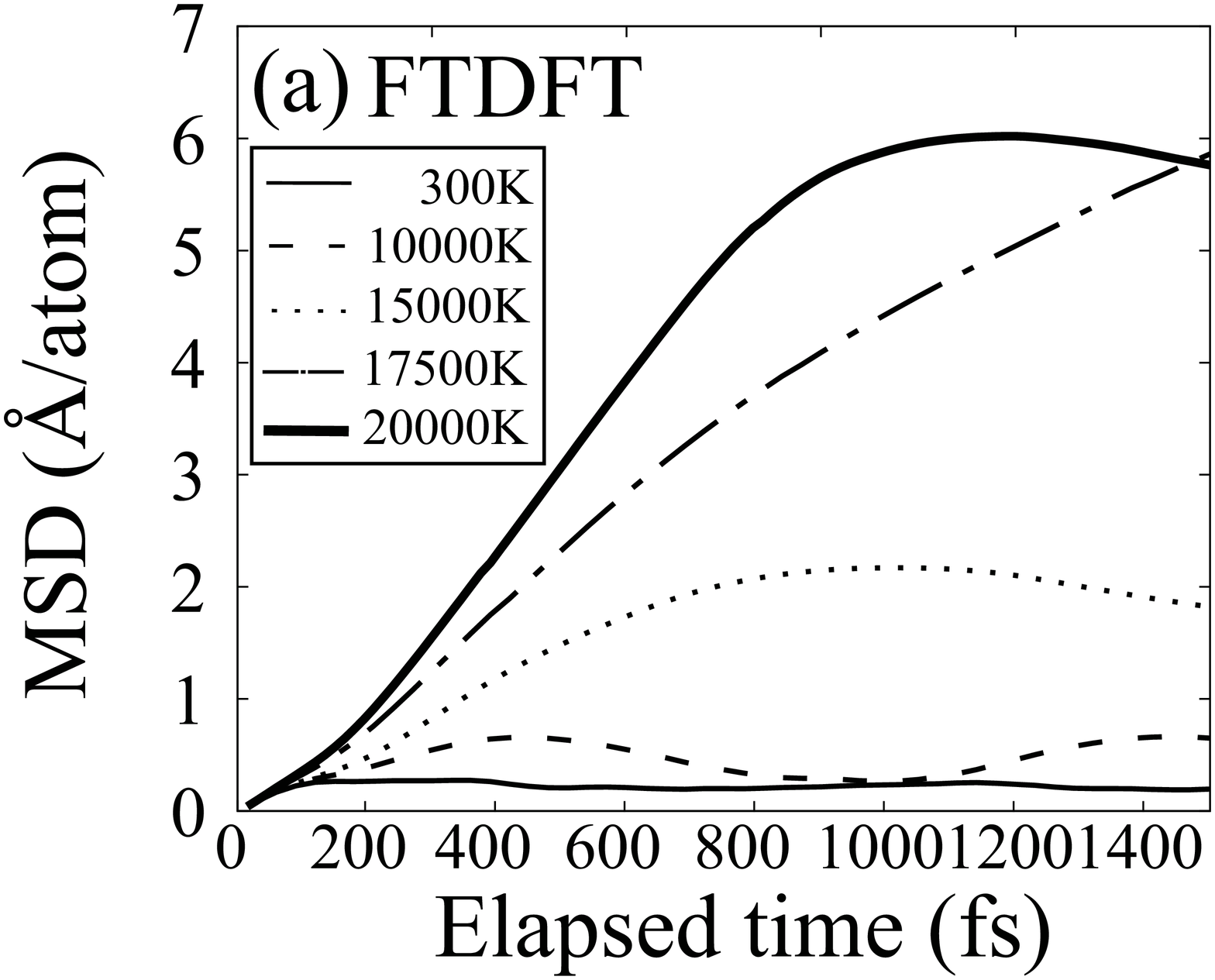}
        \end{center}
      \end{minipage} 
      \begin{minipage}{0.5\hsize}
        \begin{center}
          \includegraphics[clip, width=4.0cm]{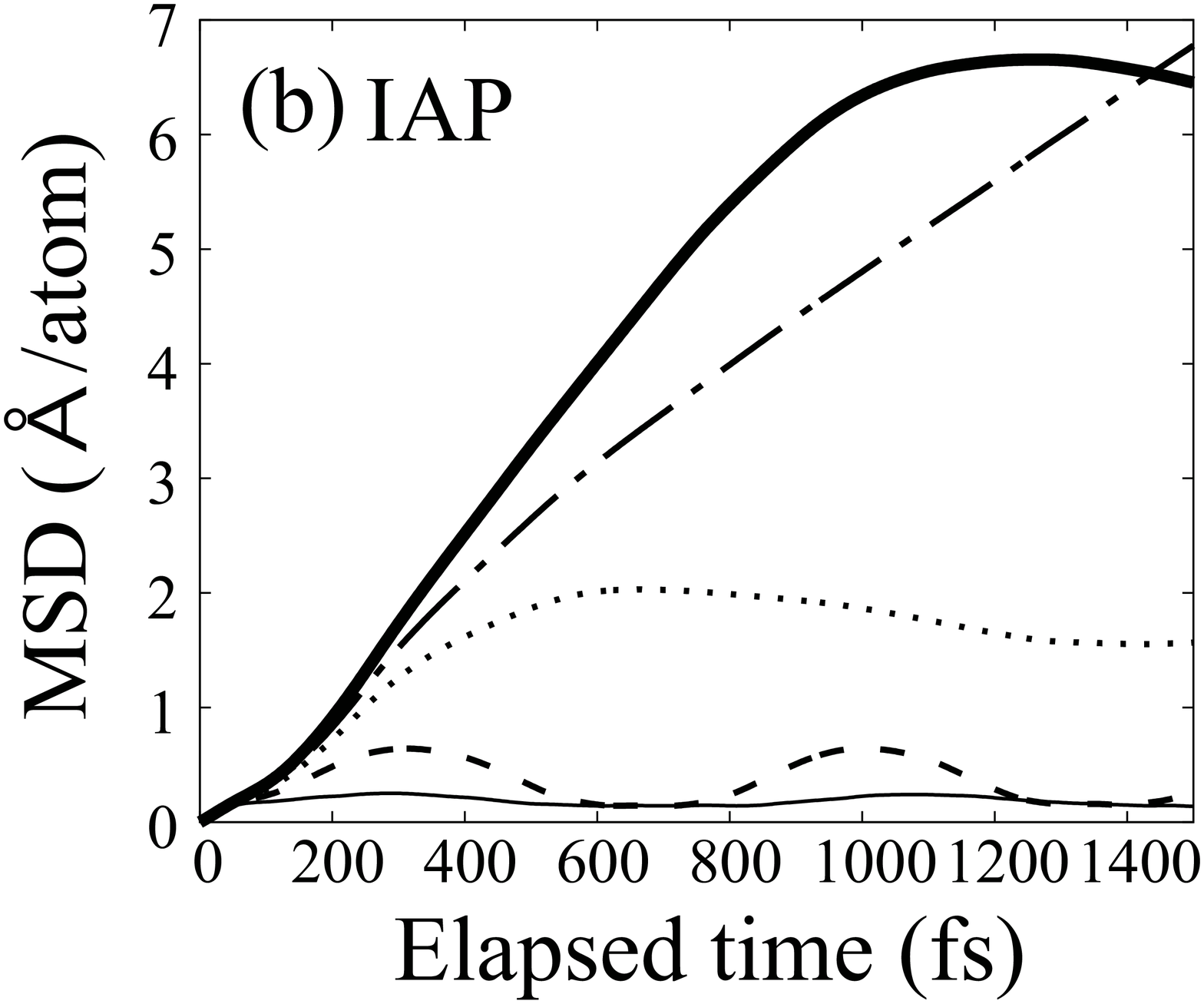}
        \end{center}
      \end{minipage} 
	
   \end{tabular}
    \label{fig:MSD_IAP}
     \end{minipage}
         \caption{
    MSD calculation results of (a) FTDFT calculations and (b) IAP calculations.
     Solid, dashed, dotted, chain, and bold lines represent $T_e=300$, $10000$, $15000$, $17500$, and $20000\,\text{K}$, respectively.
                }
        \label{fig:MSD}
\end{figure}

 Figure~\ref{fig:MSD} shows the calculation results of the MSD.
 Figure~\ref{fig:MSD}(a) shows the FTDFT calculation results, and Fig.~\ref{fig:MSD} (b) shows the IAP calculation results.
At low $T_e$, the vibration period of the IAP is shorter than that of FTDFT.
 This difference is consistent with the large phonon frequency (Fig.~\ref{fig:phonon}) and with the large value of the bulk modulus (Table~\ref{tb:lcb}). 
 Vibration is not observed at $T_e \ge 17500\,\text{K}$, which 
 indicates that ablation occurs above $T_e=17500\,\text{K}$.
 Although there are some differences between the FTDFT and IAP results, the $T_e$-dependence of these MSDs is qualitatively consistent. 
 It should be noted that we did not conduct parameter fitting at $T_e=17500\,\text{K}$, and these values were obtained by linear interpolation.
 This result demonstrates the accuracy of the $T_e$-dependent IAP even with an interpolated $T_e$.

  \subsection{Ablation threshold electron temperature: $T_e^\text{thr}$}
 \label{sec:Tethr}

 
 \begin{figure}[htbp]
  \begin{center}
    \begin{tabular}{cc}

      \begin{minipage}{0.5\hsize}
        \begin{center}
          \includegraphics[clip, width=4cm]{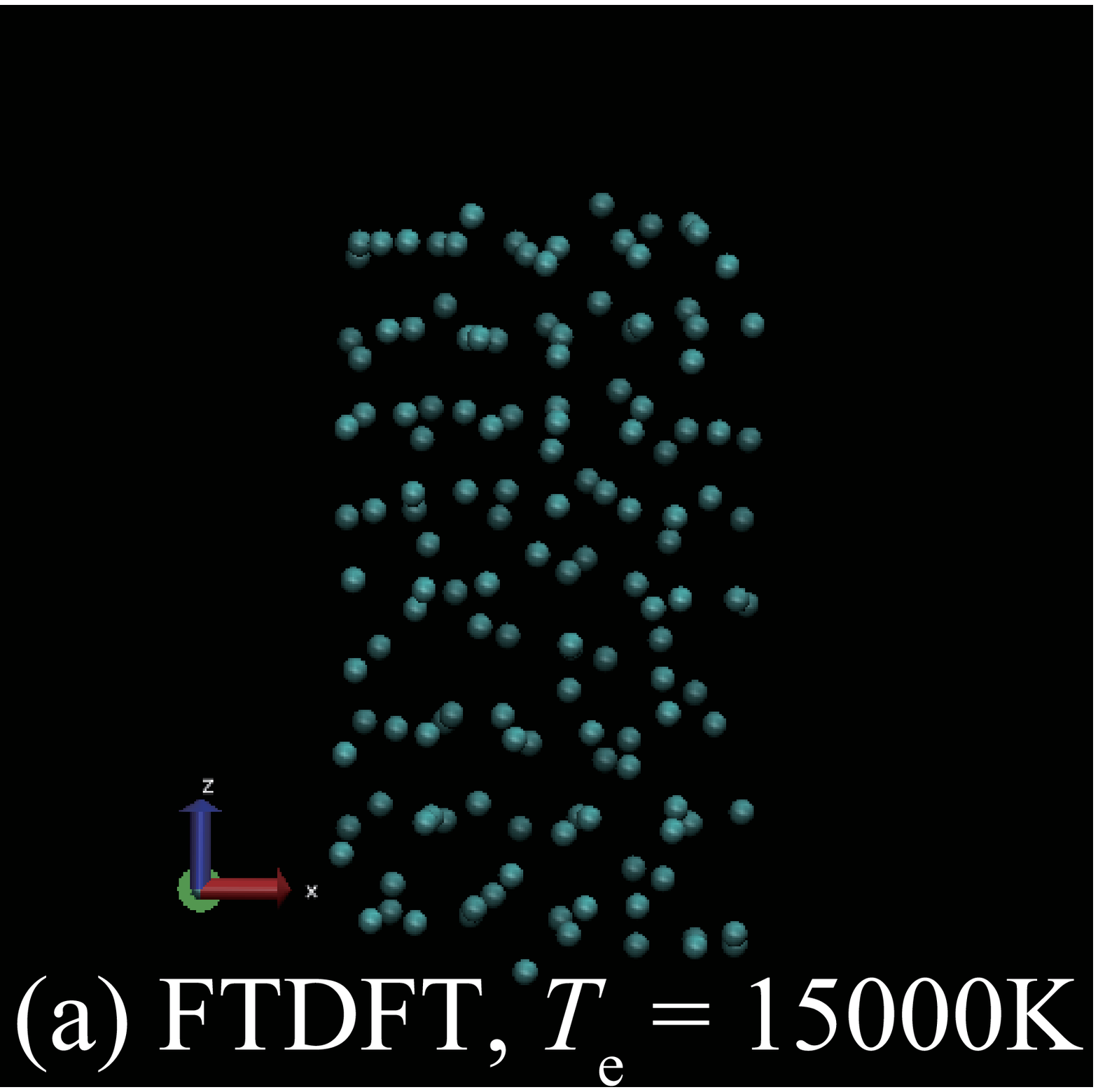}
        \end{center}  
      \end{minipage}
      \begin{minipage}{0.5\hsize}
        \begin{center}
          \includegraphics[clip, width=4cm]{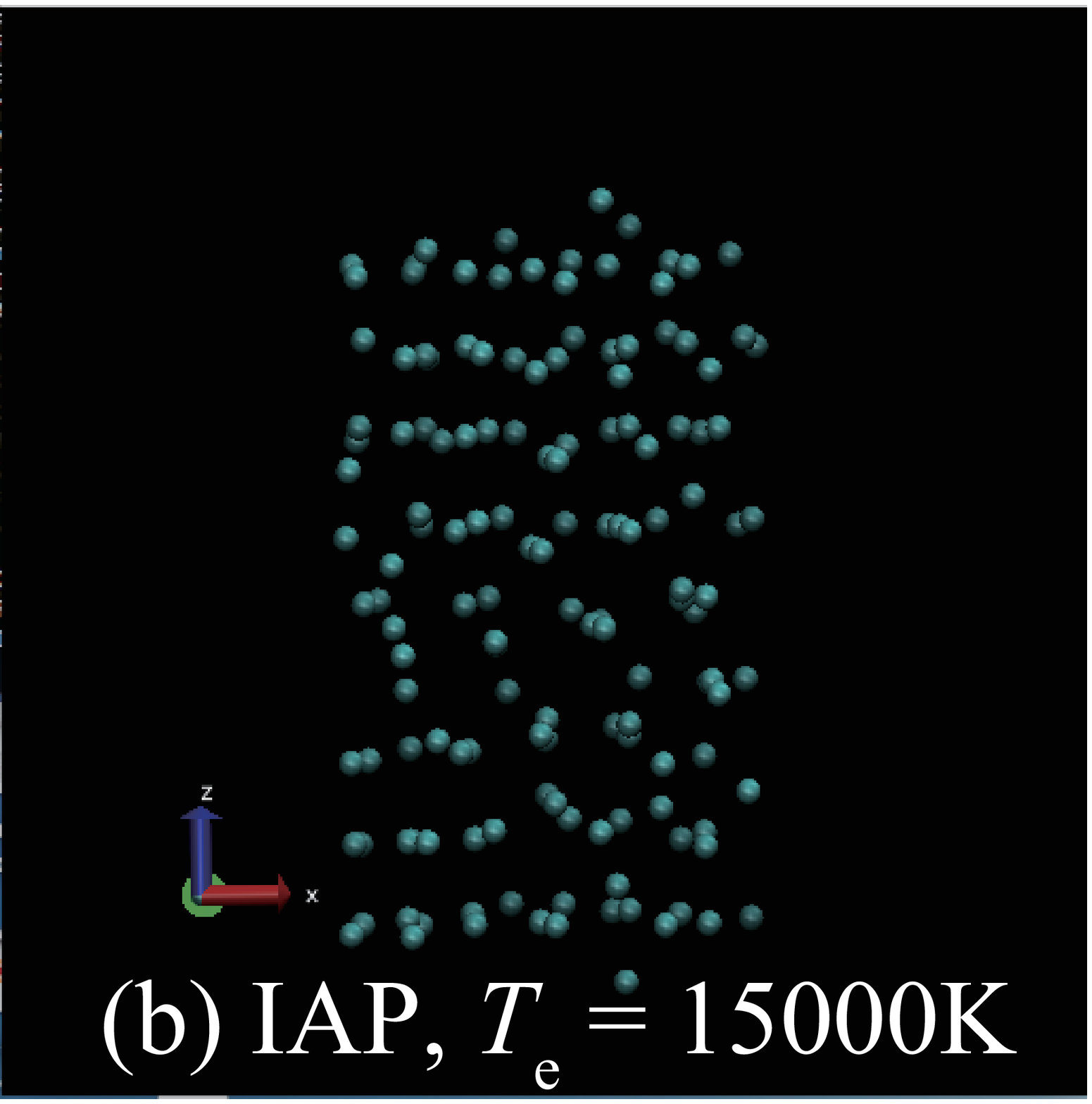}
        \end{center}
      \end{minipage}
      \\
      \begin{minipage}{0.5\hsize}
        \begin{center}
          \includegraphics[clip, width=4cm]{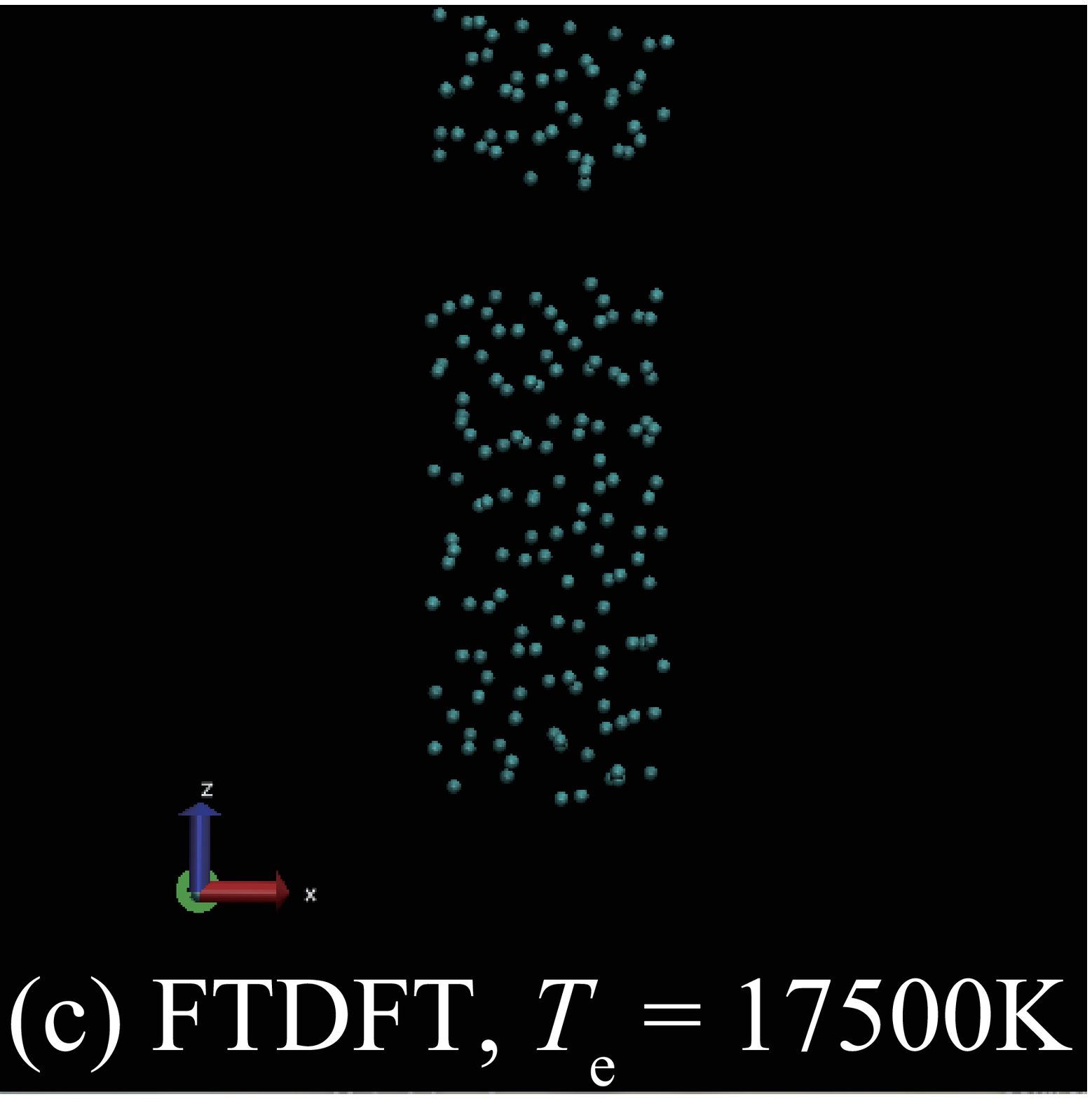}
        \end{center}
      \end{minipage}
      \begin{minipage}{0.5\hsize}
        \begin{center}
         \includegraphics[clip, width=4cm]{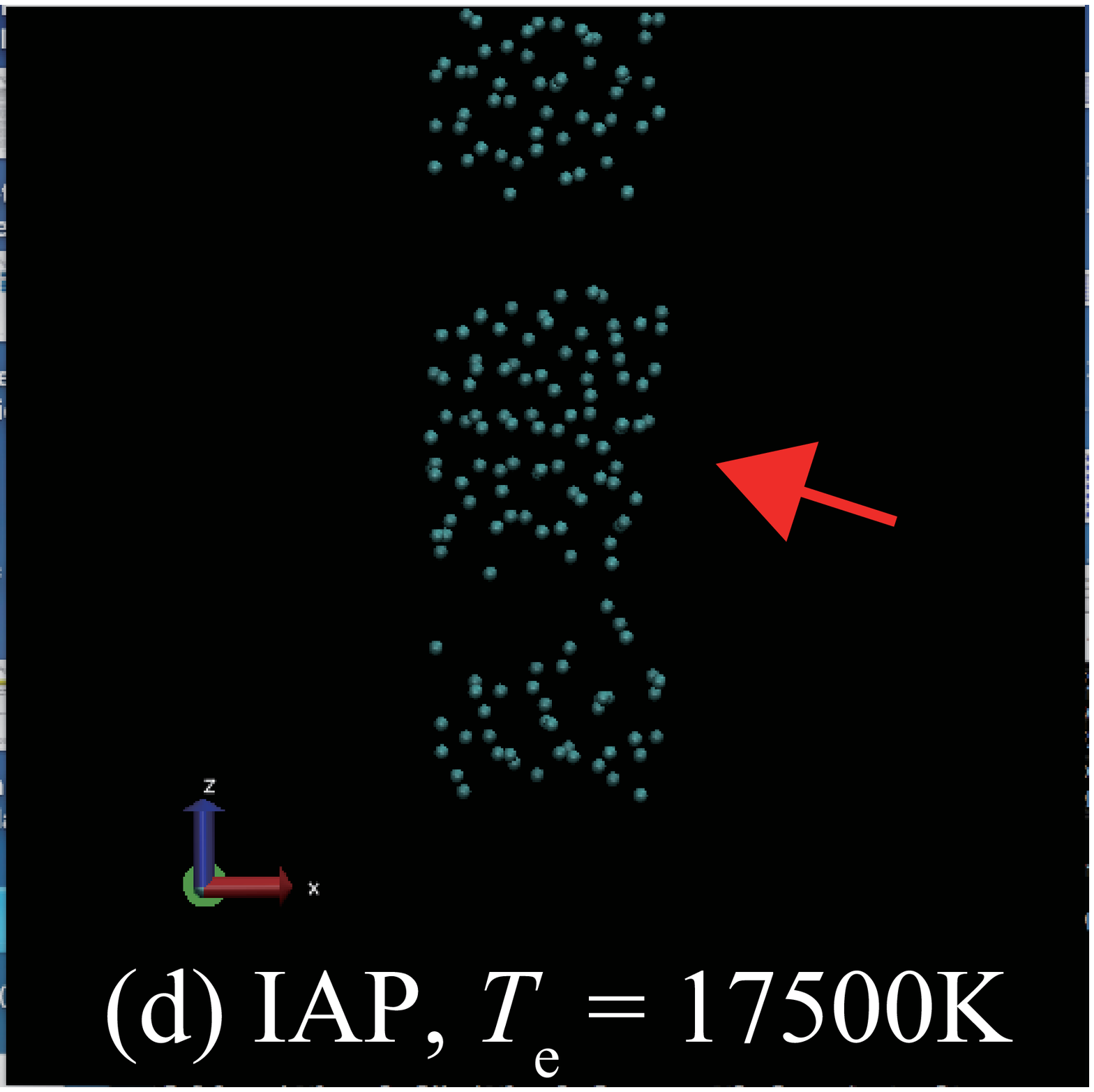}
        \end{center}
      \end{minipage}
      \\
      \begin{minipage}{0.5\hsize}
        \begin{center}
        \includegraphics[clip, width=4cm]{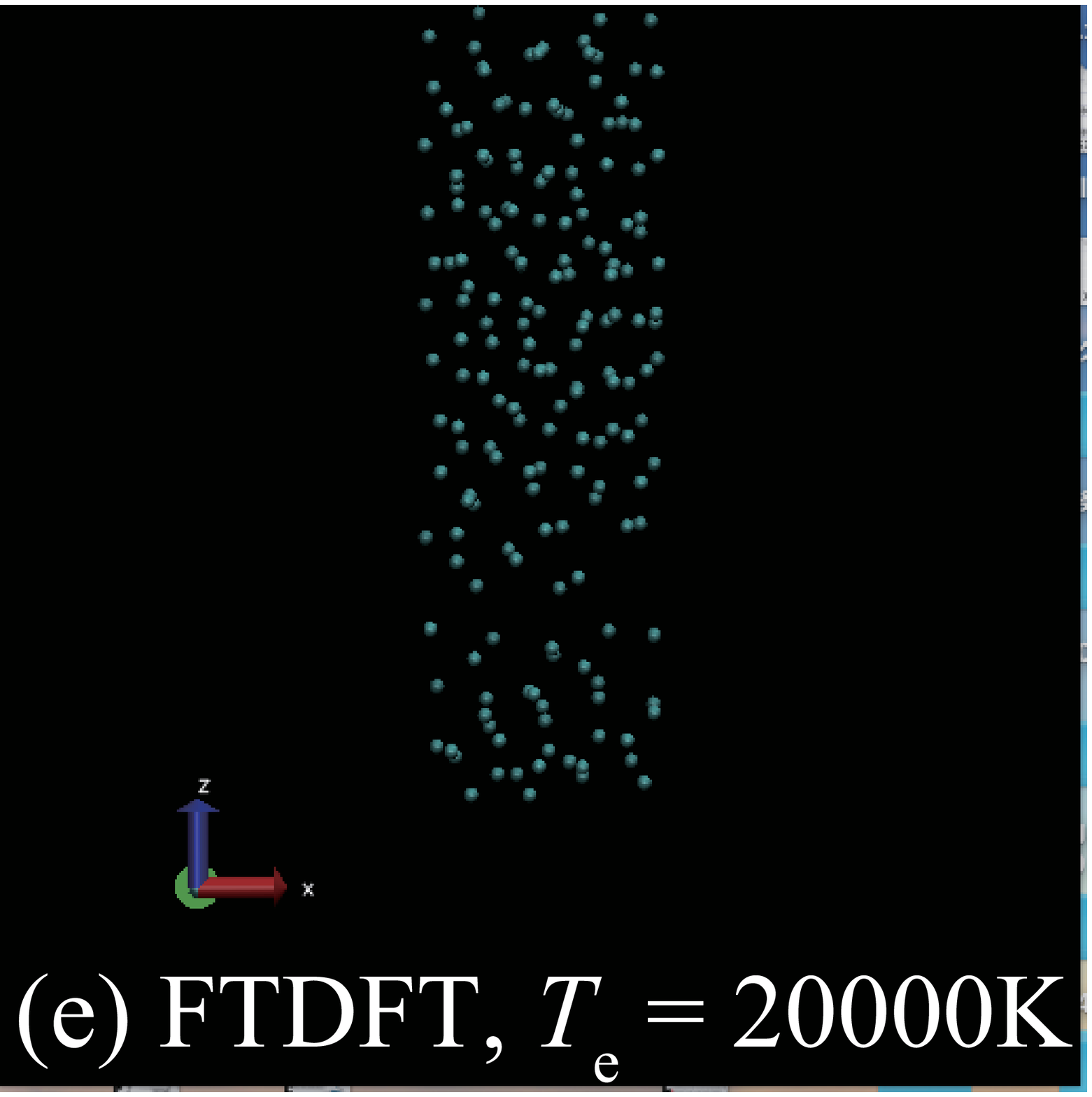}
        \end{center}
      \end{minipage}
      \begin{minipage}{0.5\hsize}
        \begin{center}
          \includegraphics[clip, width=4cm]{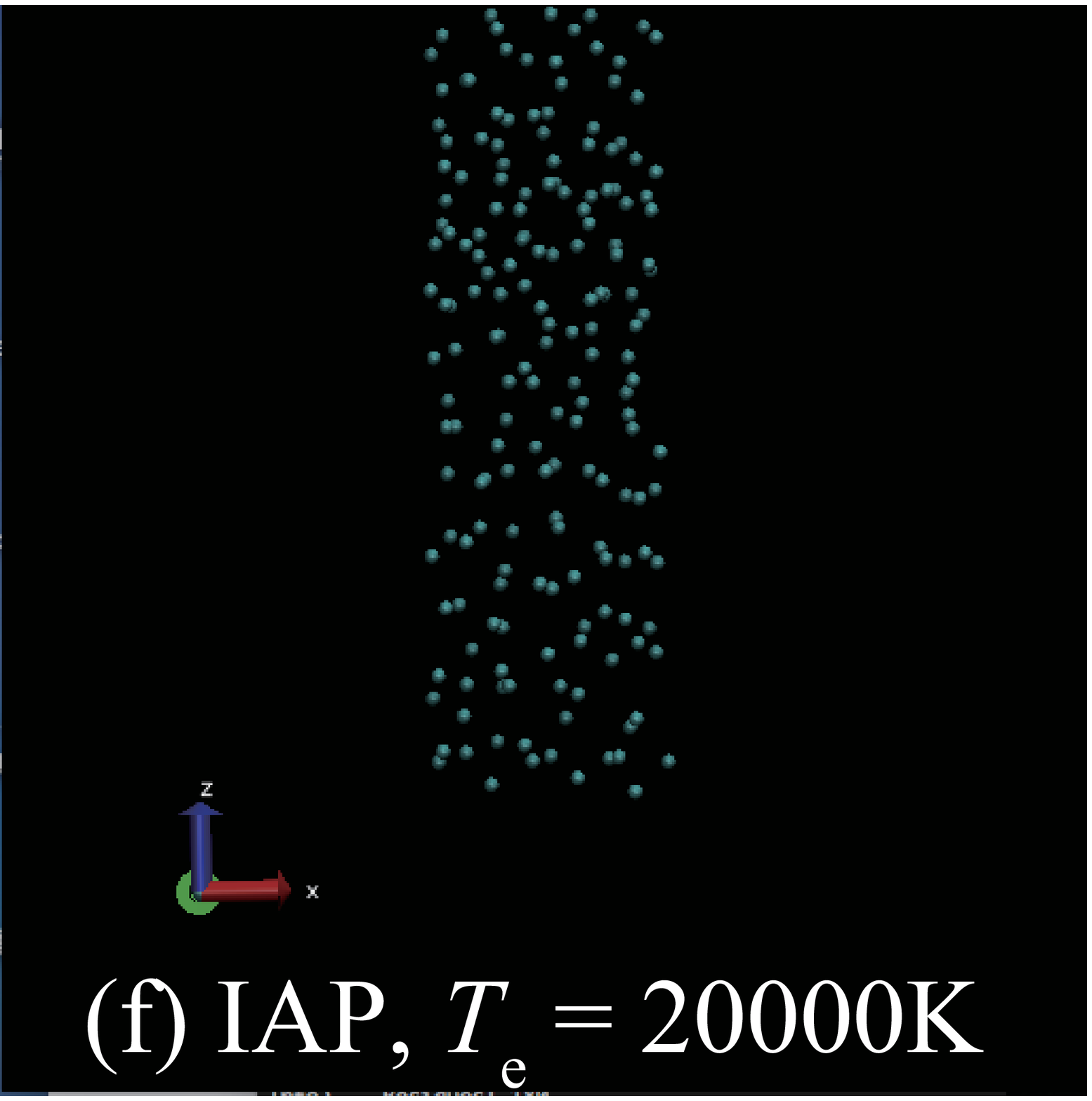}
        \end{center}
      \end{minipage}
   \end{tabular}
      \caption{    Snapshots of MD simulations for a small film system of $t=900\,\text{fs}$ after $T_e$ is changed to high $T_e$. 
               Each panel represents simulations at (a,b) $T_e=15000\,\text{K}$, (c,d) $T_e=17500\,\text{K}$, and (e,f) $T_e=20000\,\text{K}$.
               The left panels [(a), (c), (e)] show the results of first-principles MD calculations based on FTDFT, and the right panels [(b), (d), (f)] are results of classical MD simulations using the IAP. 
               These figures were visualized using Visual Molecular Dynamics (VMD).~\cite{VMD}
            }
    \label{fig:snaps}
  \end{center}
\end{figure}

Table~\ref{tb:Tethr} shows a comparison of $T_e^\text{thr}$ between the FTDFT and IAP calculations.
These numbers represent the number of times that ablation occurs at each $T_e$.
The total trial number was three.
From Table~\ref{tb:Tethr}, the discrepancy of $T_e^\text{thr}$ between the IAP results and FTDFT results is determined to be less than $500\,\text{K}$.
   
 Figure~\ref{fig:snaps} shows snapshots of MD simulations with a small film system at $900\,\text{fs}$ after $T_e$ is changed to each high $T_e$. 
 Figure~\ref{fig:snaps}(d) shows that a cluster-like material, which is indicated by the red arrow, is emitted in the IAP calculation at $T_e=17500\,\text{K}$.
 On the other hand, Fig.~\ref{fig:snaps}(c) shows that atomic-like materials are emitted in the FTDFT calculation at $T_e=17500\,\text{K}$. 
  
\begin{table}[bt]
\begin{center}
\caption{ Calculation results for $T_e^\text{thr}$.
The number represents the number of times that ablation occurs at each $T_e$.
The total trial number is three. }
    \small
      \scalebox{1}[1]{
\begin{tabular}{ccccccc} \hline 
   $T_e (10^{3}\,\text{K})$           & $15.0$  & $15.5$  & $16.0$  &  $16.5$ & $17.0$  & $17.5$   \\ \hline\hline
  IAP        & $0$ & $0$ & $0$ & $0$ & $2$ & $3$  \\ 
 FTDFT   & $0$ & $0$ & $0$  & $0$ & $3$ & $3$   \\ \hline
\end{tabular}
}

\label{tb:Tethr}
\end{center}
\end{table}

\subsection{Interpolation of $E$, $F$, and $-ST_e$}
\label{sec:interpolation}

\begin{figure}[tb]
  \begin{center}
    \begin{tabular}{c}

      \begin{minipage}{1.0\hsize}
        \begin{center}
          \includegraphics[clip, width=5.5cm]{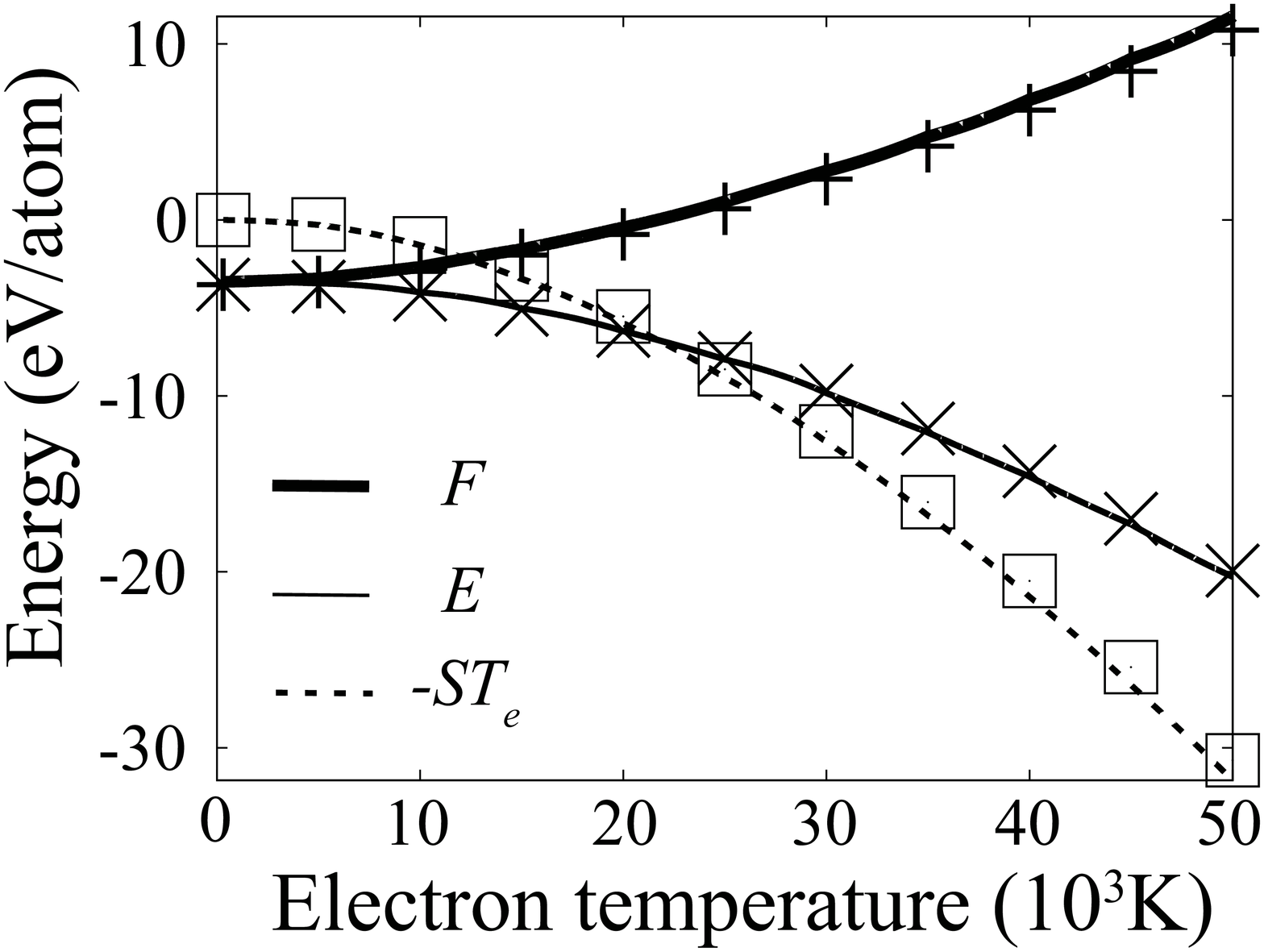}
       
        \end{center}
      \end{minipage} \\
	
   \end{tabular}
    \caption{$T_e$-dependence of $E$, $F$, and $-ST_e$.
    Solid, bold, and dashed lines represent $E$, $F$, and $-ST_e$ from the ISP calculations, respectively.
    Plus, cross, and square marks represent $E$, $F$, $-ST_e$ from the FTDFT calculations, respectively.
    The calculated structure is the equilibrium fcc structure at $T_e=300\,\text{K}$.
}
    \label{fig:Tedependent}
  \end{center}
\end{figure}

In this study, the linear interpolation method was used to determine the parameter values for $E$, $F$, and $-ST_e$.
Here, we verify the adequacy of the interpolation.
Figure~\ref{fig:Tedependent} shows the $T_e$-dependence of these values, where 
solid, bold, and dashed lines represent $E$, $F$, and $-ST_e$ calculated with the IAP, respectively, while the symbols represent the results by FTDFT.
The calculated structure is the equilibrium fcc structure.
From Fig.~\ref{fig:Tedependent}, the error due to the linear interpolation method is expected to be small in these calculations, especially at low $T_e$.
  
\begin{figure}[tb]
  \begin{center}
    \begin{tabular}{cc}

      \begin{minipage}{1.0\hsize}
        \begin{center}
          \includegraphics[clip, width=6cm]{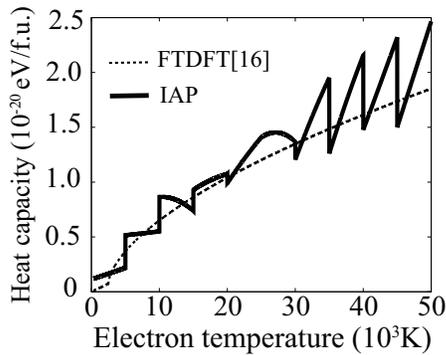}
       
        \end{center}
      \end{minipage} \\
	
   \end{tabular}
    \caption{Calculation results of $C_e(T_e)$. The dashed line represents the values estimated from FTDFT calculations.~\cite{Tanaka_2018}
    The bold line represents the IAP results.   
    The calculated structure is the fcc structure at $T_e=300\,\text{K}$.
              }
    \label{fig:heatcap}
  \end{center}
\end{figure}

\subsection{Electronic heat capacity: $C_e(T_e)$}
\label{sec:heatcap}
  
 Figure~\ref{fig:heatcap} shows the $T_e$-dependence of $C_e(T_e)$.
Figure~\ref{fig:heatcap} shows that there is a large discrepancy between the IAP and FTDFT calculations.
 The reason for this can be attributed to the linear interpolation; therefore, a spline interpolation approach can be considered to solve this problem.
  The important value for the ablation simulation is the integrated value of $C_e(T_e)$, i.e., the internal energy $E$.
 Figure~\ref{fig:Tedependent} shows the accuracy of $E$; therefore, difference of $C_e(T_e)$ is not expected to be crucial.

\section{Conclusion}
\label{sec:connclusion}

The IAP was developed for high-electron temperature simulation and the validity of the developed IAP was demonstrated.

First, we extended a formalism of the EAM potential to the $T_e$-dependent potential under the rectangular model.
We then showed a fitting methodology for parameters of the developed $T_e$-dependent IAP.
The developed $T_e$-dependent IAP was applied to copper, and its validity was demonstrated by a comparison of several physical properties, such as the energy-volume curve, phonon dispersion, electronic heat capacity, ablation threshold, and the MSD of atoms, with those calculated by FTDFT. 
The calculated results of the developed $T_e$-dependent IAP for properties important to describe laser ablation, such as the MSD and ablation threshold $T_e$, were in good agreement with the results of these properties calculated by FTDFT.
These results indicate that laser ablation caused by ultrashort pulse laser irradiation of metals can be simulated with certain accuracy.

\begin{acknowledgments}
  This work was supported in part by the Innovative Center for Coherent Photon Technology (ICCPT) in Japan and by JST COI Grant Number JPMJCE1313. 
  Y. T. was supported by the Japan Society for the Promotion of Science (JSPS) through the Program for Leading Graduate Schools (MERIT)\@.
 \end{acknowledgments}

\end{document}